\documentclass{optica-article}
\usepackage[utf8]{inputenc}
\usepackage{textgreek}
\journal{opticajournal} % for journals or Optica Open

\articletype{Research Article}

\usepackage{lineno}
\nolinenumbers % Turn off line numbering for Optica Open preprint submissions.

\begin{document}

\title{Experimental Design Space Exploration of Ultra-Low Threshold Hybrid III-V/Si Quantum Dot Microring Lasers}

\author{Xucheng Yang,\authormark{1} Preston Luong,\authormark{1} Yatiraj Ramanujam,\authormark{2} Antoine Descos,\authormark{3} Yingtao Hu,\authormark{3} Yuan Yuan,\authormark{4} Bassem Tossoun,\authormark{3} Geza Kurczveil,\authormark{3} Eunso Shin,\authormark{1} Jonathan Wierer,\authormark{1}  Ray Beausoleil,\authormark{3} Di Liang,\authormark{2} and Stanley Cheung\authormark{1}}

\address{\authormark{1}North Carolina State University, Department of Electrical and Computer Engineering, 2410 Campus Shore Dr., Raleigh, NC 27606, USA\\
\authormark{2}University of Michigan, Ann Arbor, Department of Electrical and Computer Engineering, 1301 Beal Ave., Ann Arbor, MI 48109, USA\\
\authormark{3}Hewlett Packard Labs, Large-Scale Integrated Photonics Laboratory, 820 N. McCarthy Blvd., Milpitas, CA 95035, USA\\
\authormark{4}Northeastern University, Department of Electrical Engineering, 500 MacArthur Blvd., Oakland, CA 94613, USA
}

\email{\authormark{*}scheung3@ncsu.edu} %% email address is required; see note below about the corresponding author designation

% use {asbstract*} to suppress the copyright line. Copyright information will be added in production

\begin{abstract*} 
In this work, we report on the design strategies and experimental validation of ultra-low threshold ($< 0.8\,\mathrm{mA}$) hybrid III--V/Si quantum dot (InAs/GaAs) micro-ring lasers with optical output powers $> 2\,\mathrm{mW}$ for $1.3\,\mu\mathrm{m}$ emission. The multi-dimensional design exploration allows for the demonstration of record wall-plug efficiencies ($\sim 10\%$) and threshold current densities ($109\,\mathrm{A/cm^2}$) for these compact sources on silicon. We also demonstrate the thermal performance of several designs with record characteristic temperature values of $T_0 = 212\,\mathrm{K}$, indicating minimal temperature dependence of the threshold current. In addition, the high differential gain allows for the demonstration of 3-dB bandwidths up to $5\,\mathrm{GHz}$.

\end{abstract*}

%%%%%%%%%%%%%%%%%%%%%%%%%%  body  %%%%%%%%%%%%%%%%%%%%%%%%%%
\section{Introduction}
\begin{figure}[htbp]
\centering\includegraphics[width=13cm]{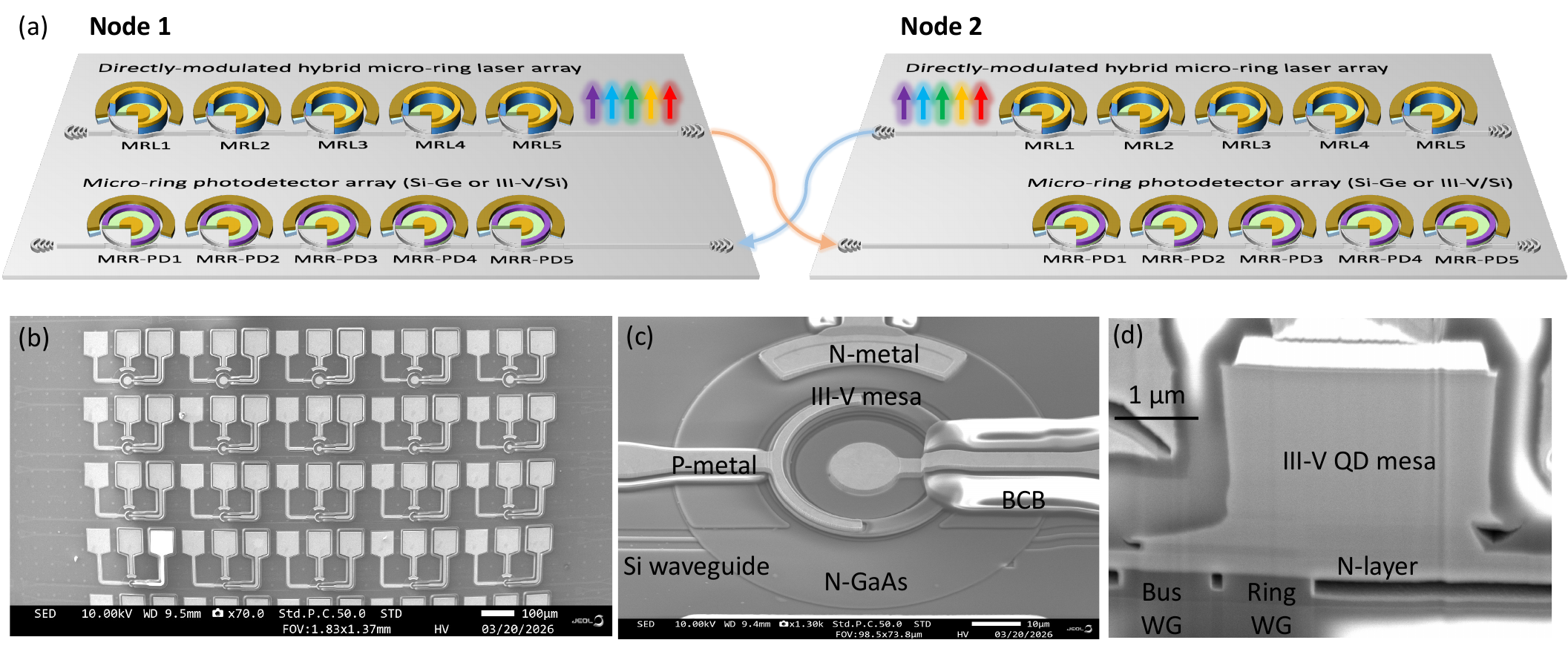}
\caption{(a) Schematic of a micro-ring based bi-directional transmit/receive (Tx/Rx) module. SEM images of (b) arrays of MRL, (c) single MRL, and (d) cross-section.}
\label{Fig1}
\end{figure}

The continued scaling of optical interconnects and integrated photonic systems has created an urgent need for compact, energy-efficient, and manufacturable on-chip light sources~\cite{Wan2026_NR}. Silicon photonics has emerged as a leading platform for large-scale integration due to its compatibility with CMOS fabrication, low propagation loss, and high index contrast enabling dense integration~\cite{Abrams2020_JLT,Cheng2018_Optica,Daudlin2025_NP}. In addition, silicon photonics development continually aims to reduce system-level power consumption to a few sub-picojoules per bit (pJ/bit), increase aggregate bandwidths to multiple terabits per second (Tb/s), and lower manufacturing costs by leveraging well-established complementary metal--oxide--semiconductor technologies~\cite{Wan2026_NR,Liang2020_OFC,Norberg2025_OFC,Piels2023_OFC}. However, silicon's indirect bandgap fundamentally limits its ability to provide efficient optical gain, necessitating the integration of III--V compound semiconductors to realize electrically pumped lasers on silicon~\cite{Liang2009_IOP}. Over the past 20 years, hybrid III--V/Si integration has therefore become a central strategy for achieving high-performance on-chip light sources~\cite{Liang2021_LAM,Li2025_OFC,Akulova2023_OFC}. In hybrid platforms, a III--V gain medium is bonded or epitaxially integrated onto a pre-patterned silicon waveguide circuit, enabling strong optical confinement in silicon while leveraging the superior gain properties of III--V materials. Compared to heterogeneous epitaxial growth approaches that rely on thick III--V buffer layers~\cite{Wan2017_Optica,Wan2018_PR}, hybrid microring architectures allow precise mode engineering, reduced material consumption, and improved compatibility with silicon foundry processes. Among the various resonator geometries explored, microring lasers offer several compelling advantages over comb lasers~\cite{Wang2017_LSA,Xian2023_OFC,Kurczveil2022_OFC,Kurczveil2020_FiO,Kurczveil2018_ISLC,Cheung2022_JLT,Cheung2022_CLEO}: 1) Precise control of channel spacing and improved tolerance to fabrication variability through local wavelength trimming~\cite{Cheung2025_NC,Cheung2024_ISLC,Tossoun2024_Arxiv}; 2) Compact footprint ($< 2000\,\mu\mathrm{m}^2$) enabling high integration density and dense wavelength-division multiplexing (DWDM) scalability~\cite{Cheung2025_NC,Liang2018_OFC,Liang2019_ACP,Cheung2022_PR}; 3) Individual direct modulation capability~\cite{Zhang2019_Optica}, and 4) Yield and redundancy through laser sparing. Fig. \ref{Fig1} illustrates one possible configuration of a DWDM bi-directional Tx/Rx module based on compact micro-ring lasers and photodetectors intended for short-distance ($30$--$50\,\mathrm{m}$) scale-up within intra-rack and inter-rack configurations~\cite{Priyadarshi2026_IEEECM,Yi2026_JSTQE,Lee2025_IEDM}.

\begin{table}[ht]
\centering
\fontsize{7pt}{7pt}\selectfont
\caption{State-of-the-Art Electrically Driven Hybrid III-V/Si Micro-Ring Lasers}
\label{tab:table1}
\begin{tabular}{lcccccccc}
\hline
Authors & Active & Wave. & Ring Radius & $I_\mathrm{th}$ & $J_\mathrm{th}$ & Peak Power & Peak & Slope \\
        & region & [$\mu$m] & [$\mu$m]   & [mA]            & [A/cm$^2$]      & [mW]       & WPE [\%] & Eff. [W/A] \\
\hline
D. Liang \cite{Liang2016_NP}          & QW  & 1.55 & 20.0  & 8.0  & 2122  & N/A  & N/A  & N/A   \\
D. Liang \cite{Liang2011_JSTQE}          & QW  & 1.55 & 25.0  & 7.62 & 3334  & 2.4  & N/A  & 0.22  \\
D. Liang \cite{Liang2009_OE}          & QW  & 1.53 & 25.0  & 8.37 & 2020  & 0.3  & N/A  & 0.025 \\
D. Liang \cite{Liang2018_OFC}          & QW  & 1.34 & 25.0  & 10.0 & 2546  & 3.3  & 4.93 & 0.1   \\
S. Cheung \cite{Cheung2025_NC,Cheung2024_ISLC,Cheung2025_OFC}   & QW  & 1.35 & 24.25 & 12.0 & 2625  & 0.3  & 0.4  & N/A   \\
T. Spuesens* \cite{Spuesens2011_G4}      & QW  & 1.55 & 7.0   & 0.6  & 389   & 0.031 & 0.5 & 0.002 \\
S. Sui \cite{Sui2015_PR}            & QW  & 1.59 & 30.0  & 4    & 997   & 0.06 & 0.17 & 0.004 \\
J. Campenhout* \cite{Campenhout2008_PTL}    & QW  & 1.59 & 7.5   & 0.9  & 509.3 & 0.012 & 0.25 & 0.008 \\
J. Campenhout* \cite{Campenhout2007_OE}    & QW  & 1.60 & 7.5   & 0.5  & 282.9 & 0.010 & 0.5  & 0.03  \\
C. Zhang \cite{Zhang2019_Optica}          & QD  & 1.31 & 25.0  & 2.0  & 254   & 0.6  & 1.2  & 0.03  \\
\textbf{This work}          & \textbf{5QD} & \textbf{1.31} & \textbf{25.0} & \textbf{0.77} & \textbf{108.9} & \textbf{1.22} & \textbf{3.54}  & \textbf{N/A}   \\
\textbf{This work}          & \textbf{5QD} & \textbf{1.31} & \textbf{25.0} & \textbf{1.17} & \textbf{165.5} & \textbf{1.61} & \textbf{9.78}  & \textbf{0.169} \\
\hline
\multicolumn{9}{l}{\footnotesize QW: quantum well; QD: quantum dot; Peak WPE: wall-plug efficiency taken at peak power} \\
\multicolumn{9}{l}{\footnotesize *Micro-disk, N/A for slope efficiency: difficult to assess due to mode-hopping} \\
\end{tabular}
\end{table}

Within the micro-ring laser (MRL), the resonant enhancement of the circulating optical power reduces the threshold carrier density, enabling low threshold currents ($< 1\,\mathrm{mA}$) and reduced energy consumption per bit ($< 1.2\,\mathrm{pJ/bit}$)~\cite{Wan2017_Optica,Wan2018_PR,Zhang2019_Optica,Liang2016_NP}. In addition, the traveling-wave nature of microring resonators mitigates spatial hole burning effects that can limit performance in Fabry--P\'erot (FP) cavities. These properties make hybrid III--V/Si microring lasers particularly attractive for short-reach data communications, optical computing, and emerging photonic co-packaging architectures~\cite{Wan2017_Optica,Wan2018_PR,Cheung2025_NC,Cheung2024_ISLC,Tossoun2024_Arxiv,Zhang2019_Optica,Liang2016_NP,Tossoun2020_IEDM,Tossoun2025_JSTQE,Tossoun2025_OFC,Lufungula2023_IPC,Cheung2025_OFC}.
These recent advances have demonstrated electrically injected hybrid microring lasers with low threshold currents, narrow linewidths, and compatibility with wafer-scale bonding processes. The incorporation of quantum well (QW) or quantum dot (QD) active regions has further improved temperature stability and reduced threshold sensitivity~\cite{Zhang2019_Optica,Liang2021_Optica}. In particular, QD gain media provide reduced linewidth enhancement factors and suppressed carrier diffusion, enabling improved coherence and potentially unidirectional operation in compact resonant geometries~\cite{Dong2021_PR,Shi2026_LSA}. Despite significant progress, several design challenges remain. Laser performance is highly sensitive to the spatial overlap between the III--V gain region and the silicon optical mode, the coupling coefficient between the microring and bus waveguide, scattering losses at bonding interfaces, and thermal management within the compact cavity. The interplay between modal confinement, material absorption, carrier injection efficiency, and resonator quality factor complicates design optimization. Achieving ultra-low threshold operation ($< 1\,\mathrm{mA}$) while maintaining high output power ($> 1\,\mathrm{mW}$), with decent wall-plug efficiencies ($\mathrm{WPE} > 10\%$) and manufacturability, therefore requires systematic exploration of device geometry and material parameters.

Despite rapid progress in heterogeneous integration, there remains a noticeable gap in the systematic design of III--V/Si QD microring lasers. Most reported demonstrations focus on proof-of-concept operation or performance benchmarking, with relatively limited exploration of how cavity geometry, coupling conditions, and fabrication limitations should be co-optimized specifically for quantum-dot gain media. This work attempts to bridge material properties, device physics, and resonator design to enable high-performance operation of these light sources on silicon. In this work, the emphasis on design parameter optimization allows for the demonstration of a record low threshold current density ($108.9\,\mathrm{A/cm^2}$), high wall-plug efficiency (WPE $\sim 10\%$), high slope efficiency ($0.17\,\mathrm{W/A}$), and high output power ($2.0\,\mathrm{mW}$). We analyze the impact of resonator geometry, coupling conditions, and gain-region configuration on the aforementioned performance metrics. Through experimental characterization and modeling, we identify the key trade-offs governing device performance and provide design guidelines for scalable, energy-efficient on-chip laser integration. Table \ref{tab:table1} summarizes the current state-of-the-art hybrid III--V/Si micro-ring lasers (MRLs), their performance metrics, and the results presented in this work.

\section{Design}

The general design of the III-V/Si QD MRL laser is a multi-dimensional optimization problem and dependent on several design variables such as: MRL ring radius ($R_r$), III-V mesa width ($w_\text{III-V}$), silicon waveguide width ($w_\text{si}$), and the coupling coefficient ($\kappa$). These variables can all affect performance parameters such as threshold current ($I_\text{th}$) and slope efficiency (SE). Sec.~\ref{sec:QDconf} will explore the dependency of the QD confinement factor $\Gamma_\text{QD}$ as a function of geometrical parameters such as $R_r$ and $w_\text{si}$. The cavity mirror loss $\alpha_m$ is determined by the coupling coefficient $\kappa$ which plays a role in both $I_\text{th}$ and SE and is discussed in Sec.~\ref{sec:CouplerDesign}. Finally, a general design strategy for minimizing $I_\text{th}$ and SE will be addressed in Sec.~\ref{sec:GenDesign}. 

\subsection{QD Confinement Factor Dependency}
\label{sec:QDconf}
\begin{figure}[htbp]
\centering\includegraphics[width=12cm]{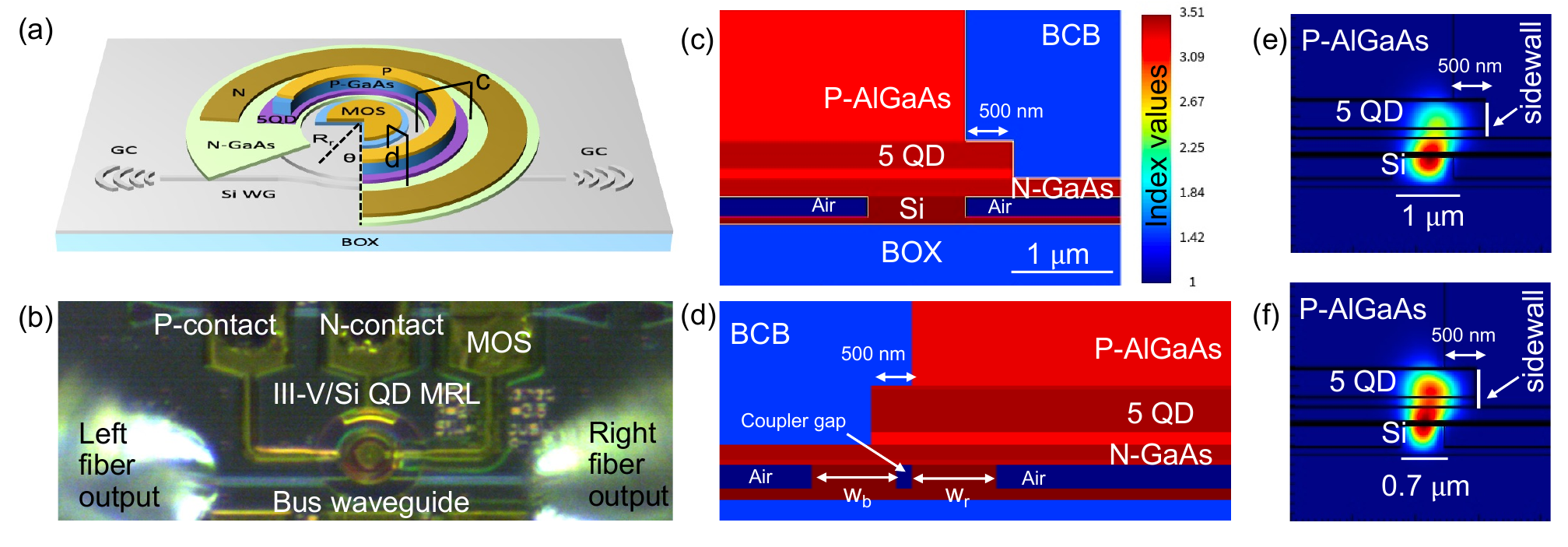}
\caption{(a) Schematic and (b) top-view image of fabricated hybrid III-V/Si 5QD MRL. Cross-sectional design and refractive index values for (c) ring mesa and (d) coupler region. Mode simulations for silicon waveguide widths of (e) $w_\text{si}$ = 1.0 \textmu m and (f) $w_\text{si}$ = 0.7 \textmu m with location of active region sidewall.}
\label{Fig2_v2}
\end{figure}

The hybrid III-V/Si QD MRL structure is defined by both the III-V epitaxial stack and silicon region as shown in Fig.~\ref{Fig2_v2}a-b. This epitaxial stack consists of 5 layers of InAs/GaAs QDs (200~nm total thickness) with a measured photoluminescence wavelength of approximately $\lambda = 1289 \pm 5$~nm. The $p$-mesa consists of $p$-AlGaAs with a thickness of 1.54~\textmu m followed by a highly doped 100~nm $p$-GaAs contact layer. The $n$-contact consists of a 150~nm thick $n$-GaAs layer with alternating superlattice layers of $n$-AlGaAs/$n$-GaAs to prevent propagation of wafer-bonding dislocations. The silicon waveguide is defined by a height and etch depth of 300~nm and 217~nm, respectively. The silicon etch depth was mainly chosen due to the etching end-point detection allowing a $\pm$10~nm variation. The cross-sectional design for the MRL mesa and coupling regions with relevant refractive index values are shown in Fig.~\ref{Fig2_v2}c-d. The 5QD active region was designed to have a 500~nm lateral offset from the edge of the GaAs mesa to minimize any possible sidewall scattering and surface carrier recombination. This is essential for MRL designs with tighter bend radii due to increased overlap of the etched sidewall if the 500~nm offset were not present. The mode calculations for $R_r = 25$~\textmu m (Fig.~\ref{Fig2_v2}e-f) also indicate the minimization of this overlap by adjusting the 5QD confinement factor ($\Gamma_\text{5QD}$) by changing the silicon waveguide widths.

\begin{figure}[htbp]
\centering\includegraphics[width=12cm]{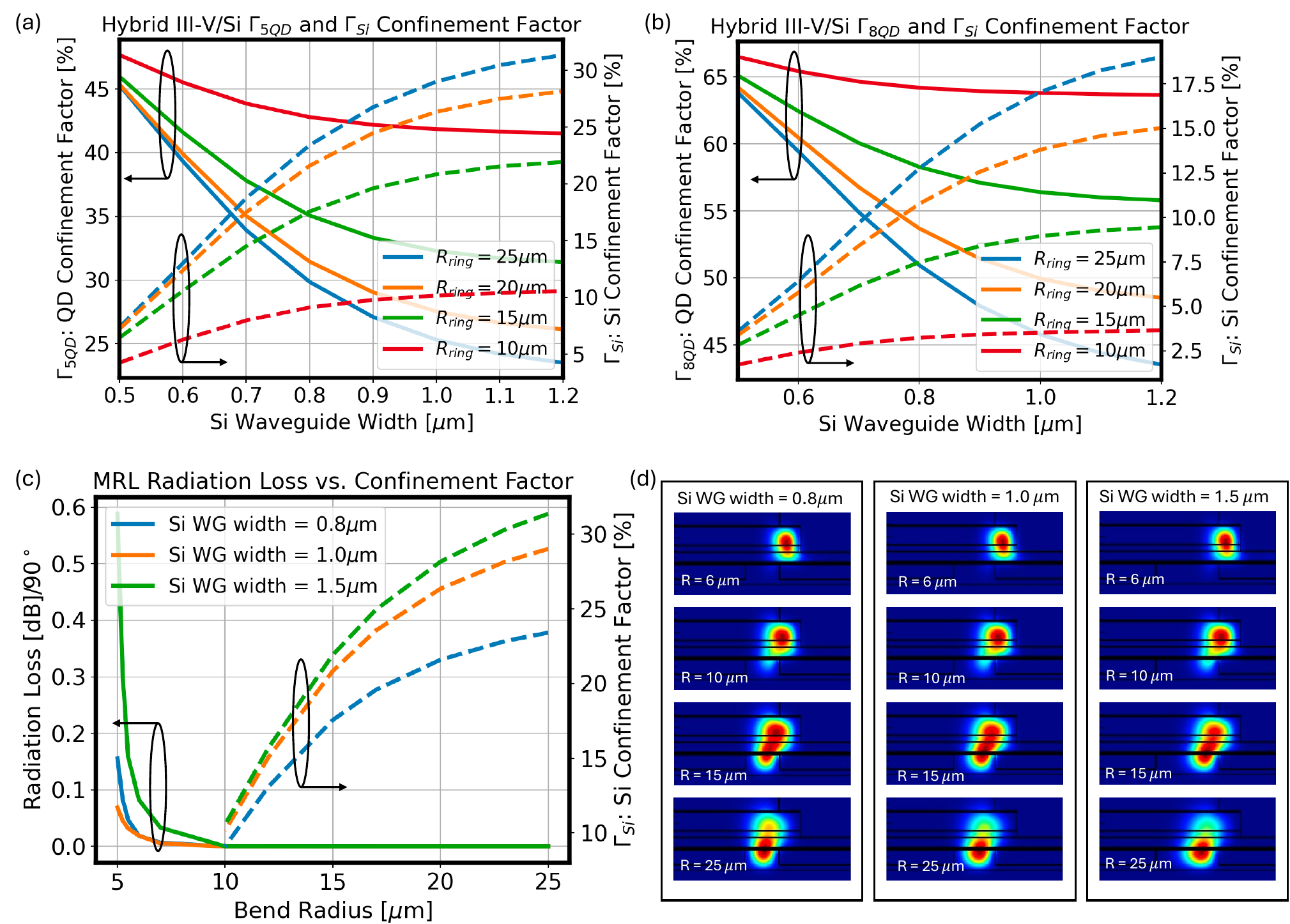}
\caption{Hybrid III-V/Si MRL QD and silicon waveguide confinement factor calculations for (a) 5QD and (b) 8QD active region for different ring radius. (c) MRL bend radiation loss vs. silicon waveguide confinement for a 5QD active region. (d) Calculated mode evolution for different silicon waveguide widths and bend radius.}
\label{Fig3}
\end{figure}

\noindent The hybrid III-V/Si MRL performance is highly dependent on both the confinement factor of the 5QD active layer ($\Gamma_\text{5QD}$) as well as the coupling coefficient ($\kappa$) between the ring and bus waveguide. Fig.~\ref{Fig3}a-b illustrates optical confinement calculations using the finite-difference eigenmode (FDE) method for both 5QD and 8QD layers, respectively. For both cases, $\Gamma_\text{QD}$ is highly dependent on silicon waveguide width and ring radius $R_r$. In general, narrower silicon waveguide widths and smaller bend radii will increase $\Gamma_\text{QD}$ at the expense of lower $\Gamma_\text{Si}$. This has the effect of lowering laser threshold currents ($I_\text{th}$) at the expense of reduced optical output power in the silicon waveguide due to minimal $\Gamma_\text{Si}$. On the other hand, too small of an $R_r$ can reduce the slope efficiency (SE) of the MRL due to increased radiation loss, as shown in Fig.~\ref{Fig3}c. Fig.~\ref{Fig3}d shows a matrix of the MRL mode evolution as a function of both silicon waveguide width and bend radius. For $R_r < 10$~\textmu m, it can be seen that the silicon modal confinement is $\Gamma_\text{Si} < 10\%$, thus preventing efficient optical mode coupling to the output silicon waveguide. Because of this, the work in this paper mainly focuses on $R_r = 10, 15, 20, 25$~\textmu m and silicon waveguide widths of $w_\text{si} = 0.8, 1.0, 1.5$~\textmu m.

\subsection{Hybrid III-V/Si Coupler Design}
\label{sec:CouplerDesign}

\begin{figure}[htbp]
\centering\includegraphics[width=13cm]{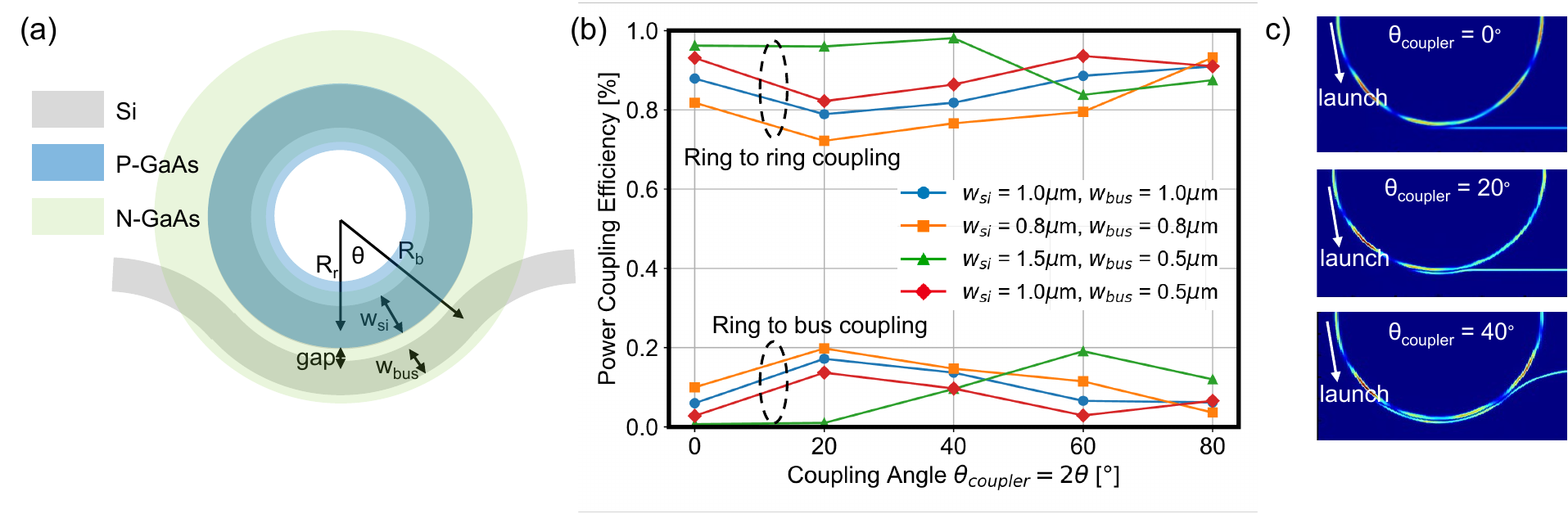}
\caption{(a) Schematic of 5QD III-V/Si ($w_\text{III-V} = 5.0$~\textmu m, $R_r = 25$~\textmu m) coupler region with relevant design parameters. (b) 3D-FDTD TE power coupling simulations for various $\theta_\text{coupler}$, $w_\text{si}$, $w_\text{bus}$ dimensions. (c) Top view of TE mode coupling for $\theta_\text{coupler} = 0, 20, 40°$.}
\label{Fig4}
\end{figure}

In order to maximize optical gain and minimize the number of active/passive transitions, the wrapped bus directional coupler region is defined by a hybrid structure as shown in Fig.~\ref{Fig4}a. The wrapped bus directional coupler was designed with phase matching such that $n_\text{eff,b} R_b = n_\text{eff,r} R_r$, where $n_\text{eff,b}$, $R_b$, $n_\text{eff,r}$, and $R_r$ are defined as the effective index of the bus waveguide, bus radius, effective index of the ring, and ring radius, respectively. The bus waveguide radius is defined as $R_b = R_r + w_r/2 + g + w_b/2$ and the ring effective index is defined as $n_\text{eff,r} = n_\text{eff,b} R_b / R_r$. Three-dimensional finite-difference time-domain (3D-FDTD) simulations indicate improved power coupling for wrapped coupler designs vs.\ point coupling ($\theta_\text{coupler} = 0°$) as shown in Fig.~\ref{Fig4}b--c. These simulations were performed for III-V mesa widths of $w_\text{III-V} = 5.0$~\textmu m for a III-V mesa radius of $R_r = 25$~\textmu m.

\noindent The ability to control the power coupling $\kappa$ enables the control of both threshold current and slope efficiency. These design parameters will be used in the following sections to determine general hybrid III-V/Si MRL designs which will then be verified experimentally.

\subsection{General Hybrid III-V/Si MRL Design}
\label{sec:GenDesign}

\begin{figure}[htbp]
\centering\includegraphics[width=13cm]{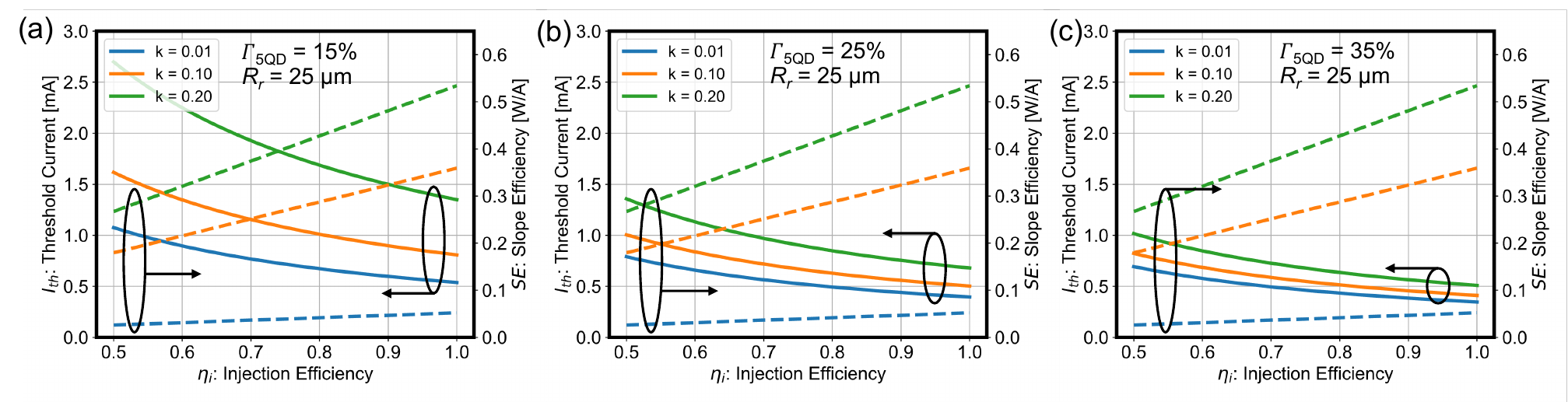}
\caption{Calculated threshold current $I_\text{th}$ and slope efficiency SE for three different power coupling coefficients $\kappa = 0.01, 0.10, 0.20$ with a MRL radius of $R_r = 25$~\textmu m assuming a confinement factor $\Gamma_\text{5QD} =$ (a) 15\%, (b) 25\%, (c) 35\%.}
\label{Fig5}
\end{figure}

The threshold current $I_\text{th}$ and slope efficiency SE for the III-V/Si QD MRL can be defined as:
\begin{equation}
    I_\text{th} = \frac{qV}{\eta_i} \left( B N_\text{tr}^2 + C N_\text{tr}^3 \, e^{\frac{\alpha_i + \alpha_\text{rad} + \alpha_m}{\Gamma_\text{QD} g_0}} \right) e^{\frac{2(\alpha_i + \alpha_\text{rad} + \alpha_m)}{\Gamma_\text{QD} g_0}}, \qquad V = \pi t_\text{QD} \left[ \left(\frac{r_2}{2}\right)^2 - \left(\frac{r_1}{2}\right)^2 \right]
    \label{eq:Ith}
\end{equation}
\begin{equation}
    \text{SE} = \eta_i \frac{\alpha_m}{\alpha_i + \alpha_\text{rad} + \alpha_m} \frac{hc}{q\lambda}, \qquad \alpha_m = \frac{1}{L_\text{ring}} \ln\!\left(\frac{1}{1-\kappa}\right)
    \label{eq:SE}
\end{equation}
where $h$, $c$, $q$, $\lambda$, $N_\text{tr}$, $\Gamma_\text{QD}$, $g_0$, and $\eta_i$ are the Planck constant, speed of light, unit electric charge, wavelength (1310~nm), transparency carrier density, quantum dot confinement factor, quantum dot material gain, and injection efficiency, respectively. The active region volume $V$ is defined by the outer and inner radius of the ring ($r_2$ and $r_1$, respectively) where $t_\text{QD}$ is the total thickness of the 5QD active region (280~nm in this work). $\alpha_i$, $\alpha_\text{rad}$, $\alpha_m$, $L_\text{ring}$, and $\kappa$ are the internal active region loss, ring radiation loss, mirror loss, total ring length, and power coupling coefficient. $\alpha_i$ in a 5QD InAs/GaAs system was reported to be $\alpha_i = 22$~cm$^{-1}$~\cite{Amano2006_APL} and will be used in this work. The bimolecular recombination coefficient $B$ and transparency carrier density $N_\text{tr}$ were taken to be $B = 1.1 \times 10^{-10}$~cm$^3$/s and $N_\text{tr} = 1.0 \times 10^{18}$~cm$^{-3}$, respectively, according to work done on subthreshold characterization of a 7QD system~\cite{Zenari2023_ACS}. The Auger recombination coefficient $C$ is a significant source of nonradiative recombination and has reported values ranging from $C = 4 \times 10^{-29}$--$8 \times 10^{-29}$~cm$^6$/s for temperatures from $T = 100$--$300$~K~\cite{Ghosh2001_APL}. The modal gain for the 5QD region was reported to be $\Gamma_\text{5QD} g_0 = 43$~cm$^{-1}$~\cite{Amano2006_APL} for a dot density of $8.0 \times 10^{10}$~cm$^{-2}$/layer and Innolume reported $\Gamma_\text{7QD} g_0 = 45$~cm$^{-1}$ with $\Gamma_\text{7QD} = 9\%$ for a 7QD region~\cite{Maximov2008_SCT,Uvin2018_OE}. Based on these reported values, it is reasonable to estimate a material gain $g_0 \sim 500$~cm$^{-1}$. As a result, the threshold current $I_\text{th}$ and slope efficiency SE can be determined as a function of injection efficiency $\eta_i$, power coupling coefficient $\kappa$, and confinement factor $\Gamma_\text{5QD}$ as shown in Fig.~\ref{Fig5}a-c. Given a constant MRL radius of $R_r = 25$~\textmu m and a fixed $\Gamma_\text{5QD}$, the threshold current $I_\text{th}$ decreases monotonically with increasing $\eta_i$ because of improved carrier injection into the active region. A value of $\eta_i = 0.87$ has been reported in several works~\cite{Zenari2023_ACS,Jung2018_ACS}. Sub-mA thresholds of $I_\text{th} < 1$~mA are possible by improving either the modal gain through $\Gamma_\text{5QD}$ or a higher quality ($Q$) factor ring via a smaller power coupling coefficient $\kappa$, albeit at the expense of lower SE. In addition, a reduction of $R_r$ by 40\% can improve the reduction of $I_\text{th}$ by 33--50\% for increasing values of $\eta_i$. All calculations for Fig.~\ref{Fig5}a-c are performed at $T = 25$~°C and neglect self-heating but serve as a general guide for designing sub-mA threshold III-V/Si QD MRLs.

\begin{figure}[htbp]
\centering\includegraphics[width=14cm]{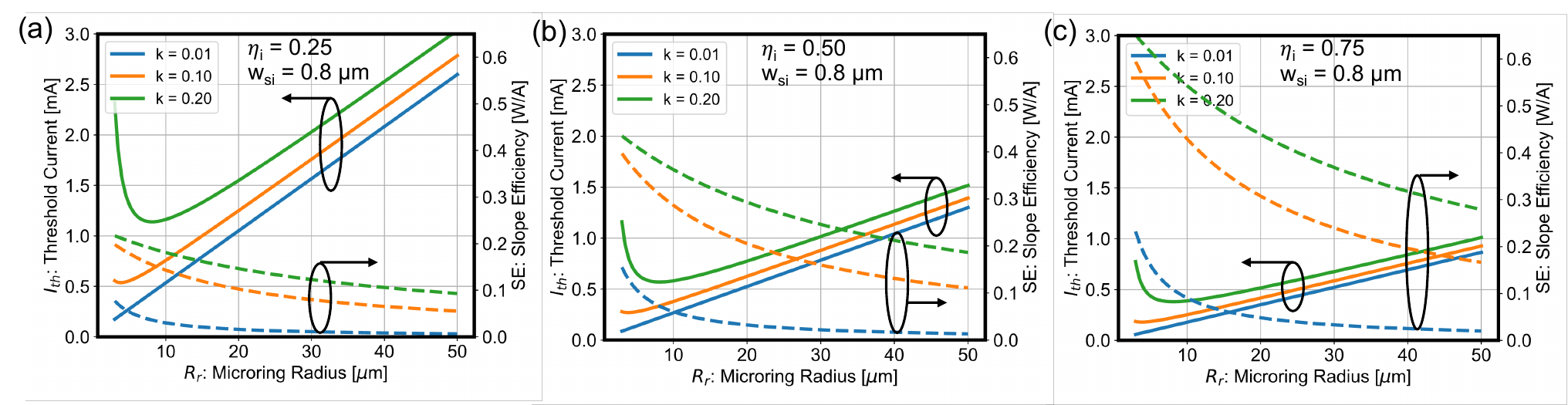}
\caption{Calculated threshold current $I_\text{th}$ and slope efficiency SE for three different power coupling coefficients $\kappa = 0.01, 0.10, 0.20$ with varying MRL radius $R_r$ assuming injection efficiencies of $\eta_i =$ (a) 0.25, (b) 0.50, (c) 0.75.}
\label{Fig6}
\end{figure}

The expected improvement for smaller MRLs with weaker coupling coefficients is shown in Fig.~\ref{Fig6}a--c. The calculations include the $\Gamma_\text{5QD}$ dependency on $R_r$ for a constant silicon waveguide width of $w_\text{si} = 0.8$~\textmu m. In fact, $\Gamma_\text{5QD}$ is dependent on both $R_r$ and silicon waveguide width $w_\text{si}$ as shown in Fig.~\ref{Fig3} and is accounted for in the calculations in Fig.~\ref{Fig5}a--c. It should also be noted that MRL devices with $R_r < 10$~\textmu m will have the majority of optical power trapped in the III-V QD region as shown in Fig.~\ref{Fig3}d and as a result have much lower extracted optical power in the silicon bus waveguide. Fig.~\ref{Fig6}c illustrates the case where injection efficiencies $\eta_i > 0.75$ can result in sub-mA threshold currents with SE $> 0.3$ for a wide range of coupling coefficient values. In reality, there is a trade-off between the desire to minimize $I_\text{th}$ by minimizing MRL device volume and the need to maintain reasonable current densities and thermal impedances. As a result, we restricted the fabricated bend radius to $R_r \geq 10$~\textmu m. It is also interesting to note that MRL designs with $\kappa = 0.20$ have an optimal $I_\text{th}$ for a particular $R_r$, mainly due to the counterbalancing of both coupler and ring radiation loss for increased $\Gamma_\text{5QD}$ at small radii. The modeling approach outline here will be used to fit experimental and characterization results detailed in Section~\ref{sec:Measurements}.

\section{Heterogeneous III-V/Si Fabrication}

In-house device fabrication starts with a 100~mm SOI wafer that consists of a 300~nm top silicon layer and a 2~\textmu m thick buried oxide (BOX) layer. A blanket ion implantation of boron was performed ($4 \times 10^{16}$~cm$^{-3}$) to facilitate conductivity of integrated volatile and non-volatile semiconductor-insulator-semiconductor capacitive (SISCAP) phase shifters. Alignment marks and grating couplers were both patterned using a 248~nm KrF ASML DUV stepper and etched 145~nm with Cl$_2$-based gas chemistry. A 9-step boron implantation scheme was used to create $p^{++}$ contacts ($1 \times 10^{20}$~cm$^{-3}$) for the SISCAP phase shifters. Next, silicon rib waveguides and vertical outgassing channels (VOCs) were patterned and etched 217~nm and 300~nm, respectively, with laser end-point detection. The silicon wafer is then thoroughly cleaned with a Piranha solution followed by a short dilute hydrofluoric (HF) acid etch to remove any residual hard masks. An O$_2$ plasma clean was then performed followed by an SC1 and SC2 clean. The III-V QD epitaxial wafer is cleansed with acetone, methanol, and IPA, followed by an O$_2$ plasma clean and a NH$_4$OH:H$_2$O (1:10) dip for 1~min. Next, a dielectric layer of Al$_2$O$_3$/HfO$_2$/Al$_2$O$_3$ is deposited on both the III-V and SOI wafer via atomic layer deposition (ALD) at 300~°C. The two samples are then mated manually at room temperature using a Finetech flip-chip bonder and then wafer-bonded under pressure at 300~°C (2~hour ramp) for a total of 15~hours. 

\begin{figure}[htbp]
\centering\includegraphics[width=13cm]{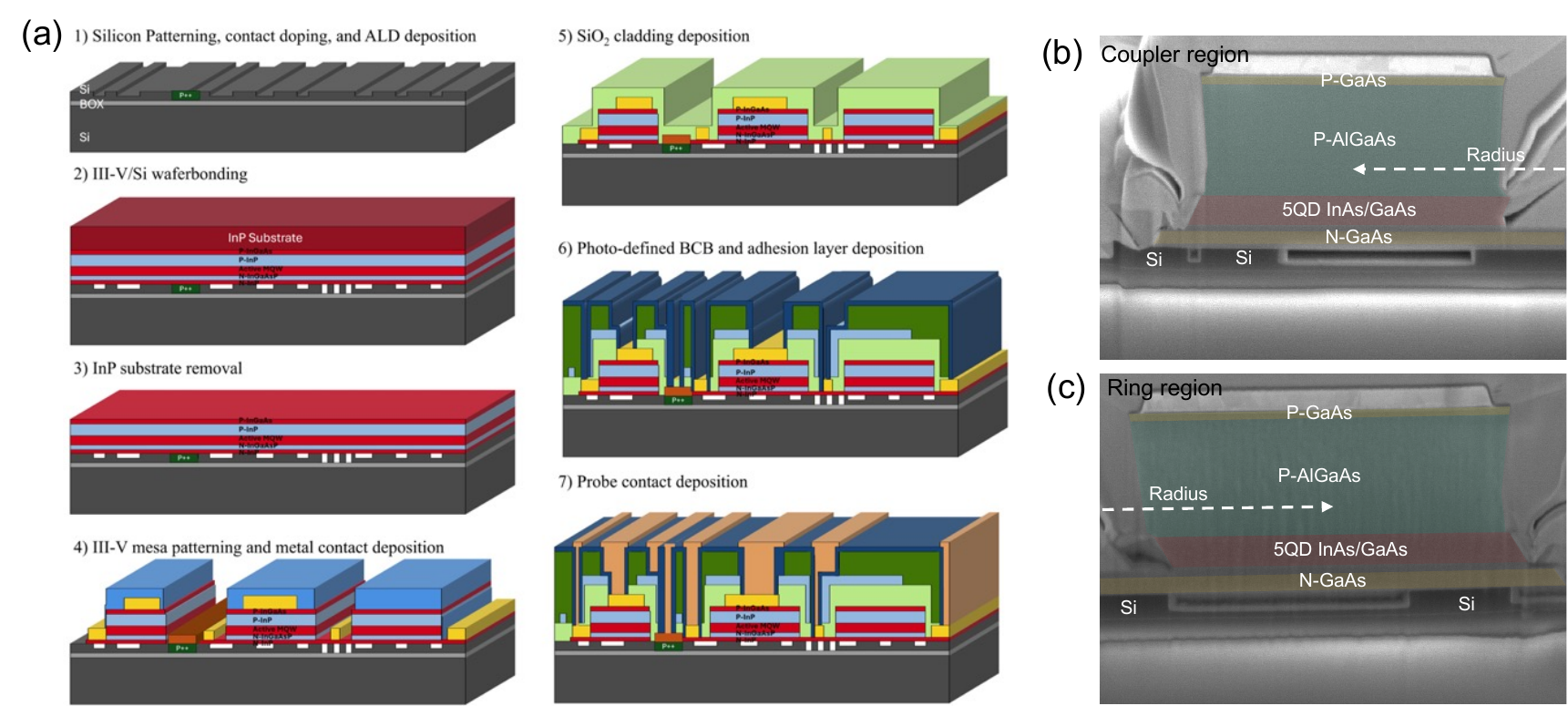}
\caption{(a) Detailed fabrication flow of hybrid III-V/Si QD MRL devices, (b)-(c) cross-sectional SEM images of coupler and ring region respectively.}
\label{Fig7}
\end{figure}

After wafer-bonding, the backside of the III-V was mechanically lapped until a 100~\textmu m thickness of III-V remained. Next, the $p$-GaAs substrate is removed using a wet etch which selectively stops on a 20~nm $p$-AlGaAs layer. The etch stop layer is then subsequently removed using buffered HF acid to reveal a clean 100~nm $p$-GaAs contact layer. Metal contacts consisting of Pt/Ti/Pt/Au (5/25/50/250~nm) were deposited onto the $p$-GaAs as the laser $p$-contact metal. III-V mesas are then defined by etching using a SiN hard mask and an ICP etcher using a Cl$_2$-based gas chemistry stopping within a 100~nm $n$-AlGaAs etch stop layer using laser endpoint detection. This etch stop layer is then wet etched to reveal a clean $n$-GaAs contact. A combination of Pd/Ge/Ti/Au/Ti (30/60/50/200/10~nm) metals were deposited to create the laser $n$-contact metal and annealed at 300~°C for 30~seconds. Next, the III-V QD mesas were isolated by selectively dry etching various regions of the $n$-GaAs and ALD dielectric. Next, a plasma-enhanced chemical vapor deposition (PECVD) SiN cladding was deposited followed by a thick BCB layer to minimize electrical parasitics. Finally, a thick layer of Ti/Au (1.6~\textmu m) was evaporated to create metal probe pads for both laser $p$- and $n$-contacts.

\section{Characterization and Measurements}
\label{sec:Measurements}

\subsection{Measurement Preliminaries and Design of Experiment (DOE)}

The light-current-voltage (LIV) curves were characterized by current injection with a Keithley 2400 source measurement unit and optical power from the left and right grating coupler was collected using cleaved SMF-28 fibers positioned at an angle of 7°. Optical power from both grating couplers was measured with a Newport 2936-C power meter and optical spectra were simultaneously collected using a 90/10 directional coupler into a Yokogawa AQ6370E optical spectrum analyzer. Measured grating coupler losses for TE polarization were determined to be on average ${\sim}{-8.7}$~dB/coupler at 1310~nm. The pre-bonded grating coupler loss is ${\sim}{-7}$~dB/coupler and indicates the entire III-V/Si bonding process incurs an extra loss of 1.7~dB/coupler. The 100~mm wafer is vacuum mounted onto a copper chuck with temperature control held at $T = 25$~°C. All measured LI responses are normalized to the grating coupler losses. Pre-bonded silicon waveguides with a width of 0.5~\textmu m were measured to have TE losses of ${\sim}{-22.1}$~dB/cm at a wavelength of 1310~nm, determined from a series of spiral waveguide cut-back test structures. 

\begin{figure}[htbp]
\centering\includegraphics[width=13cm]{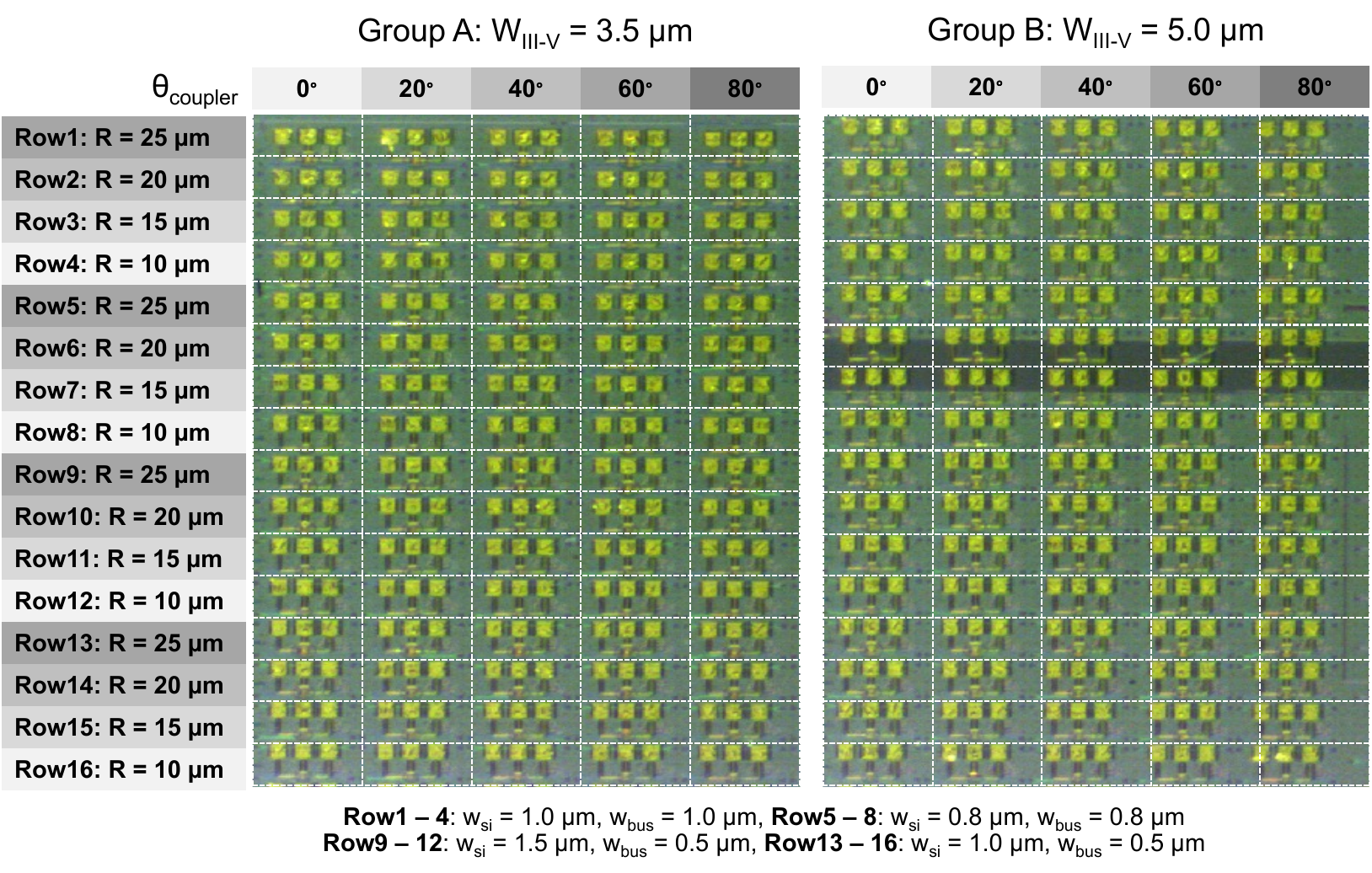}
\caption{III-V/Si QD MRL design of experiment to explore $I_\text{th}$ and SE dependency on ring radius $R_r$, coupling angle $\theta_\text{coupler}$, mesa width $w_\text{III-V}$, silicon waveguide ring width $w_\text{si}$, and silicon waveguide bus width $w_\text{bus}$.}
\label{Fig8}
\end{figure}

The fabricated design of experiments (DOE) consists of two groups A and B, where the III-V mesa width $w_\text{III-V} = 3.5$~\textmu m and 5.0~\textmu m, respectively. As shown in Fig.~\ref{Fig8}, each design group consists of variations in coupling angle $\theta_\text{coupler} = 0°, 20°, 40°, 60°, 80 ^\circ$ and ring radius $R_r = 10, 15, 20, 25$~\textmu m. Furthermore, each row consists of various directional coupler designs based on silicon ring waveguide width $w_\text{si}$ and silicon bus waveguide width $w_\text{bus}$ as indicated in the description below Fig.~\ref{Fig8}. The intention of trying symmetric and asymmetric directional couplers with various waveguide widths was to best phase-match curved coupler designs as described in Sec.~\ref{sec:CouplerDesign}.

\subsection{Experimental Characterization of Low-Threshold III-V/Si QD MRLs}

\begin{figure}[htbp]
\centering\includegraphics[width=13cm]{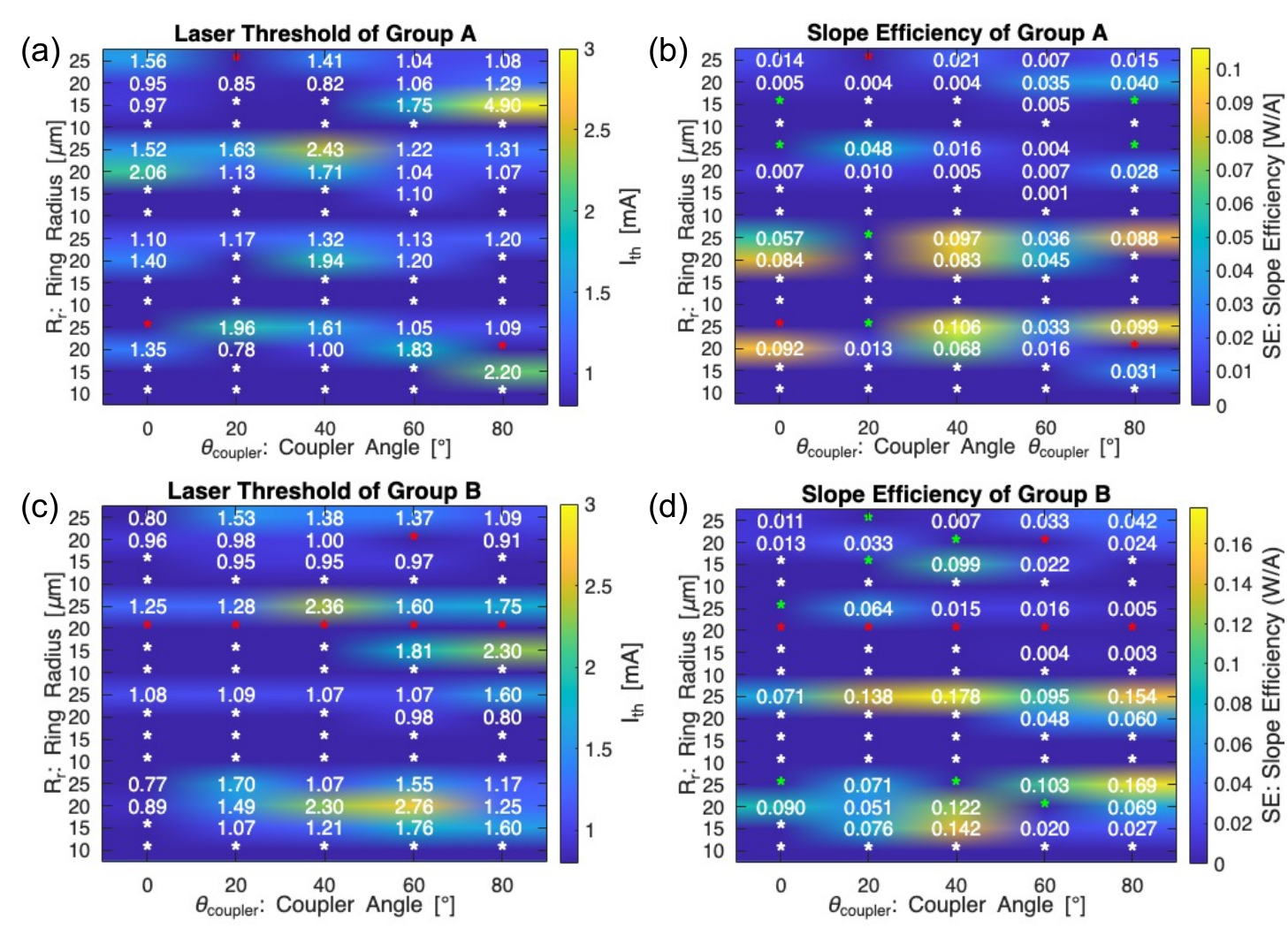}
\caption{(a)--(d) Measured $I_\text{th}$ and SE based on group A devices with $w_\text{III-V} = 3.5$~\textmu m (a)--(b) and group B devices with $w_\text{III-V} = 5.0$~\textmu m (c)--(d). (\textcolor{red}{$*$}): Absence of IV curve due to fabrication error. (\textcolor{gray}{$*$}): IV curve is present but low output optical power due to incomplete optical transition to silicon bus waveguide. (\textcolor{green}{$*$}): SE difficult to interpret due to significant mode-hopping.}
\label{Fig9}
\end{figure}

The measured lasing threshold $I_\text{th}$ and slope efficiency SE of group A and B devices are shown in Fig.~\ref{Fig9}a--b and Fig.~\ref{Fig9}c--d, respectively. As expected, the majority of devices with radius $R_r = 10, 15$~\textmu m failed to exhibit any reasonable optical output power, mainly due to the high confinement factor in the III-V QD region as shown in Fig.~\ref{Fig3}d. As a result, there is lower extracted optical power in the silicon bus waveguide. The red stars (\textcolor{red}{$*$}) indicate the lack of an observable LIV curve due to possible fabrication errors, whereas the white stars (\textcolor{gray}{$*$}) indicate observable IV curves but low optical output powers as discussed in Sec.~\ref{sec:QDconf}. Green stars (\textcolor{green}{$*$}) in the 2D maps of the slope efficiency SE (Fig.~\ref{Fig9}b, d) indicate uninterpretable SE because of significant LI instability due to mode-hopping. The threshold current $I_\text{th}$ is properly extracted by fitting a high-order polynomial with highly sampled points near threshold and by taking the 2$^\text{nd}$ derivative of the LI curve according to Telcordia Technologies standards (GR-3013-CORE)~\cite{Telcordia_GR468,Woodham2006,ILXLightwave_AN12}. 

The peak of this 2$^\text{nd}$ derivative identifies the location of $I_\text{th}$ such that reliable extraction of 160 III-V/Si MRL lasers can be made without "eyeball" guessing. Details of this method are described in the Supplementary section. Group A ($w_\text{III-V} = 3.5$~\textmu m) and B ($w_\text{III-V} = 5.0$~\textmu m) lasers exhibited an average $I_\text{th} = 1.43$~mA and 1.34~mA, respectively. The differences were determined to be inconsistencies in the coupling coefficient $\kappa$ as well as lower injection efficiency $\eta_i$ for the narrower $w_\text{III-V}$, as will be discussed later.

\begin{figure}[htbp]
\centering\includegraphics[width=13cm]{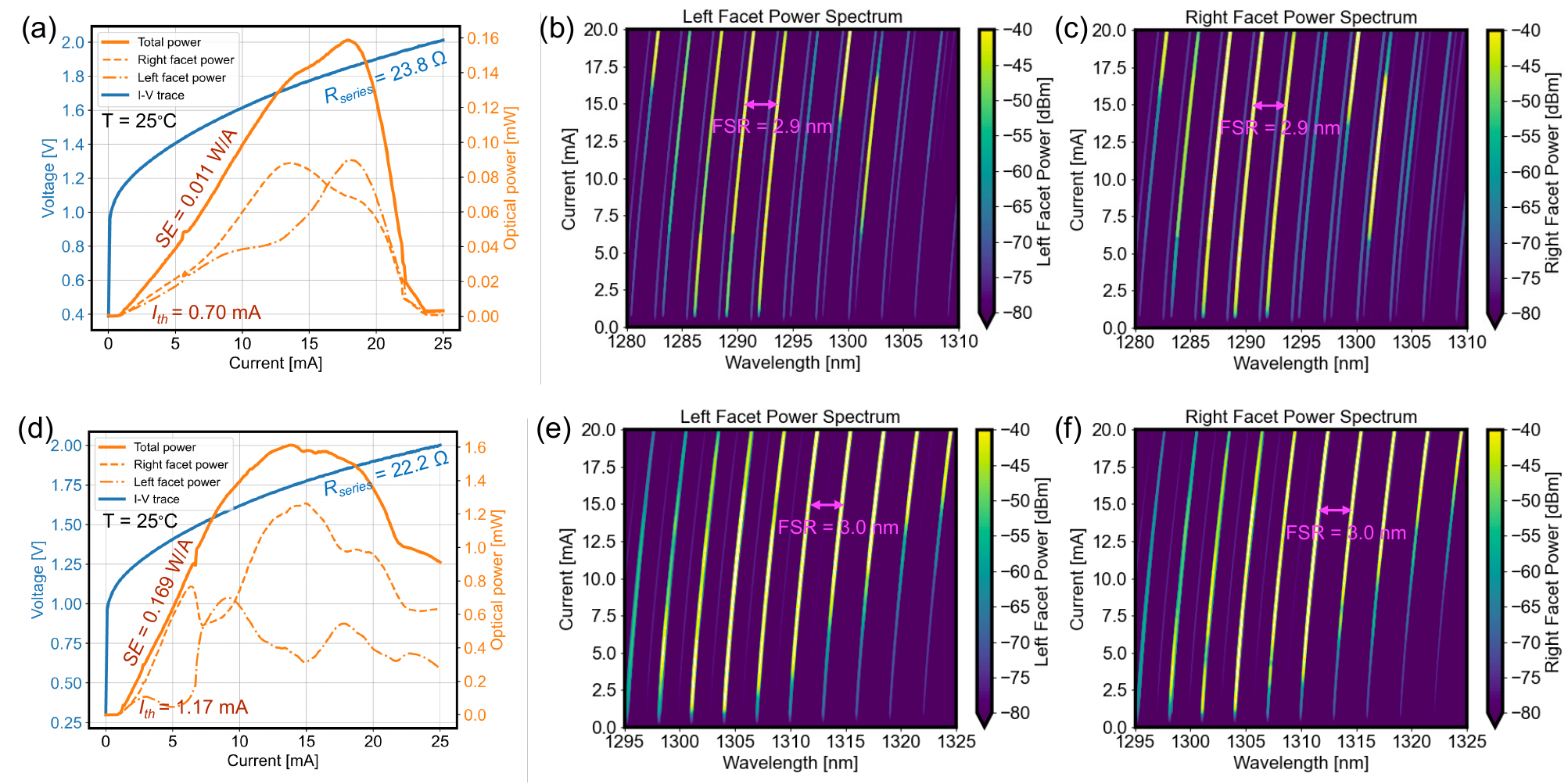}
\caption{III-V/Si MRL design ($R_r = 25$~\textmu m, $w_\text{III-V} = 5$~\textmu m, $w_\text{si} = 1.0$~\textmu m, $w_\text{bus} = 1.0$~\textmu m) with measured (a) LIV and (b)--(c) CCW/CW optical spectra. III-V/Si MRL design ($R_r = 25$~\textmu m, $w_\text{III-V} = 5$~\textmu m, $w_\text{si} = 1.0$~\textmu m, $w_\text{bus} = 0.5$~\textmu m) with measured (d) LIV and (e)--(f) CCW/CW optical spectra.}
\label{Fig10}
\end{figure}

Fig.~\ref{Fig10} shows LIV and optical spectrum measurements for two III-V/Si 5QD MRLs ($R_r = 25$~\textmu m, $w_\text{III-V} = 5.0$~\textmu m) from Group B with different coupling coefficients. Fig.~\ref{Fig10}a--c consists of a MRL design ($R_r = 25$~\textmu m, $w_\text{III-V} = 5$~\textmu m, $w_\text{si} = 1.0$~\textmu m, $w_\text{bus} = 1.0$~\textmu m, $\theta_\text{coupler} = 0°$) with a designed point power coupling coefficient of $\kappa = 0.007$, which leads to a low slope efficiency SE $= 0.011$~W/A albeit with a sub-mA threshold current of $I_\text{th} = 0.8$~mA. Fig.~\ref{Fig10}d--f consists of a MRL design ($R_r = 25$~\textmu m, $w_\text{III-V} = 5$~\textmu m, $w_\text{si} = 1.0$~\textmu m, $w_\text{bus} = 0.5$~\textmu m, $\theta_\text{coupler} = 80°$) with a designed point power coupling coefficient of $\kappa = 0.066$, which leads to a higher slope efficiency SE $= 0.169$~W/A while maintaining a slightly higher threshold current of $I_\text{th} = 1.17$~mA. At $\lambda = 1310$~nm, the measured free-spectral range is approximately $\Delta\lambda_\text{FSR} = 3.0$~nm. The experimentally determined effective group index can then be calculated using $n_g = \lambda^2 / (\Delta\lambda_\text{FSR} / 2\pi R) = 3.64$, which is close to the FDE-calculated value of $n_g = 3.55$. The slope efficiency of both lasers is determined by summing both clockwise (CW) and counter-clockwise (CCW) lasing powers. It should be noted that the LI curves for both MRLs exhibit the absence of any kinks, indicating that lasing does not occur through sequential turn-on of independent clockwise (CW) and counter-clockwise (CCW) traveling-wave modes.

In an ideal microring cavity these modes are frequency-degenerate, but in practical devices residual sidewall roughness and fabrication imperfections introduce weak backscattering that can couple the two counter-propagating directions. This coupling lifts the degeneracy and forms standing-wave super-modes that share the same optical frequency and threshold condition. As a result, lasing occurs in a coupled mode rather than in two independent traveling-wave modes, and the gain clamps once at threshold without a secondary transition as current increases. Because both directional components participate in the same lasing eigenmode, no redistribution of carrier population occurs with increasing drive current, and the output power therefore increases smoothly with injection. Consequently, the LI curve remains linear above threshold without the slope changes or kinks that would otherwise arise from the onset of an additional lasing mode or directional switching.

\begin{figure}[htbp]
\centering\includegraphics[width=13cm]{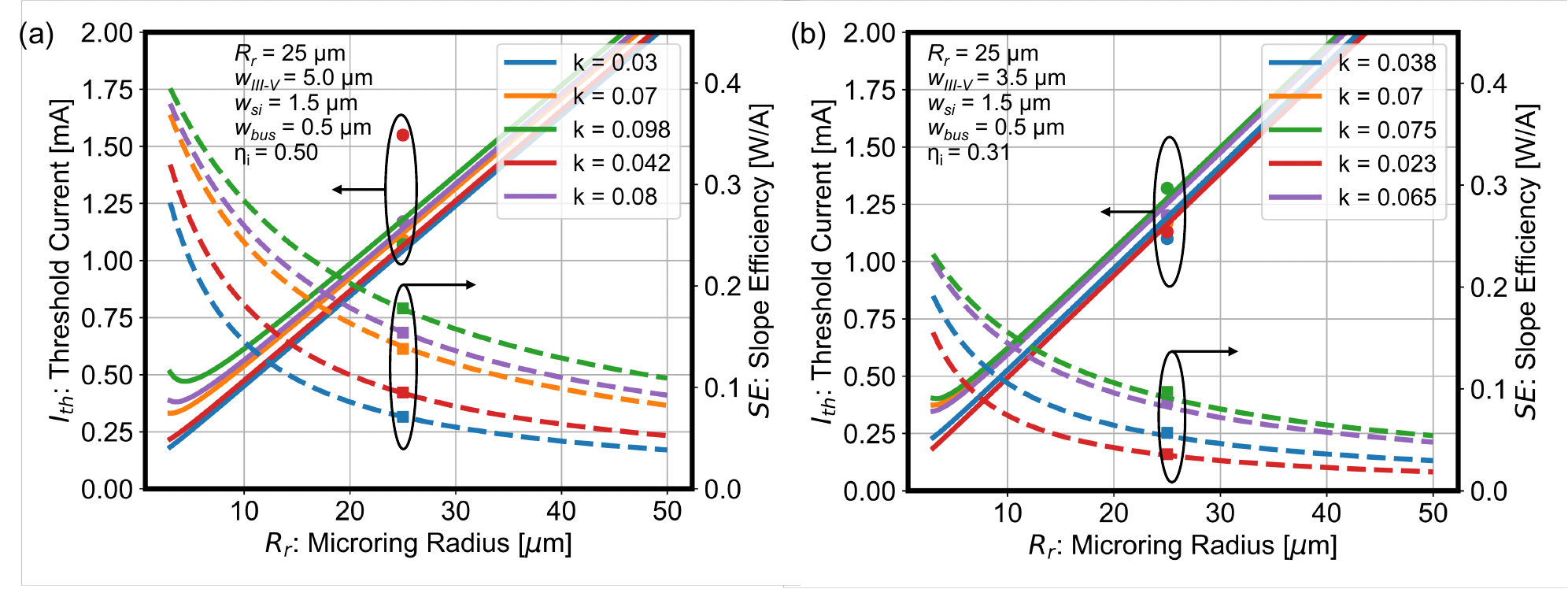}
\caption{Comparison of experimental and calculated threshold current and slope efficiency for (a) $w_\text{III-V} = 5.0$~\textmu m and (b) $w_\text{III-V} = 3.5$~\textmu m.}
\label{Fig11}
\end{figure}

Fig.~\ref{Fig11} shows a comparison of experimental and calculated threshold currents and slope efficiency for similar MRLs with the exception of (a) $w_\text{III-V} = 3.5$~\textmu m and (b) $w_\text{III-V} = 5.0$~\textmu m. The dimensions for both MRLs are $R_r = 25$~\textmu m, $w_\text{si} = 1.0$~\textmu m, $w_\text{bus} = 0.5$~\textmu m. Several material parameters were provided by the epitaxial wafer vendor (QDLaser Inc.) such as $g_0 = 3000$~cm$^{-1}$, $N_\text{tr} = 1.2 \times 10^{18}$~cm$^{-3}$, $B = 1 \times 10^{-10}$~cm$^3$/s, and $C = 3 \times 10^{-29}$~cm$^6$/s. The internal loss was assumed to be $\alpha_i = 22$~cm$^{-1}$ based on reports for a 5QD InAs/GaAs system~\cite{Amano2006_APL}. As a result, the power coupling coefficient $\kappa$ and injection efficiency $\eta_i$ play the largest role in determining the best fit of $I_\text{th}$ and SE defined by Eqs.~(\ref{eq:Ith})--(\ref{eq:SE}). In order to fit the MRL laser in Fig.~\ref{Fig11}a, we iteratively fit both the coupling coefficient and the injection efficiency for the best match to the experimental results. The power coupling coefficients were simulated to be $\kappa = 0.007, 0.01, 0.096, 0.191$, and $0.12$ for coupling angles $\theta_\text{coupler} = 0, 20, 40, 60$, and $80^\circ$. However, the fitting suggests $\kappa = 0.03, 0.07, 0.098, 0.242$, and $0.08$. The injection efficiency was best determined to be $\eta_i = 0.50$, leading to relatively close fitting of $I_\text{th}$. Fitting was also performed for the other MRL as shown in Fig.~\ref{Fig11}b, and the coupling coefficients were determined to be $\kappa = 0.038, 0.07, 0.075, 0.023$, and $0.065$ with an injection efficiency of $\eta_i = 0.31$. The difference in coupling coefficients could possibly be due to spatial fabrication errors, as these two MRLs are 3.35~mm apart. In addition, there could be variations in the III-V processing where the etched mesa in the coupler region can vary in overlap with the bus waveguide and cause differences in power coupling coefficients. The injection efficiency is also 38\% lower and may indicate surface recombination for the narrow III-V mesa. Despite these variations, the general design outlined in Sec.~\ref{sec:GenDesign} allows us to design MRLs with near sub-mA thresholds and slope efficiencies approaching 0.20~W/A.

\begin{figure}[htbp]
\centering\includegraphics[width=13cm]{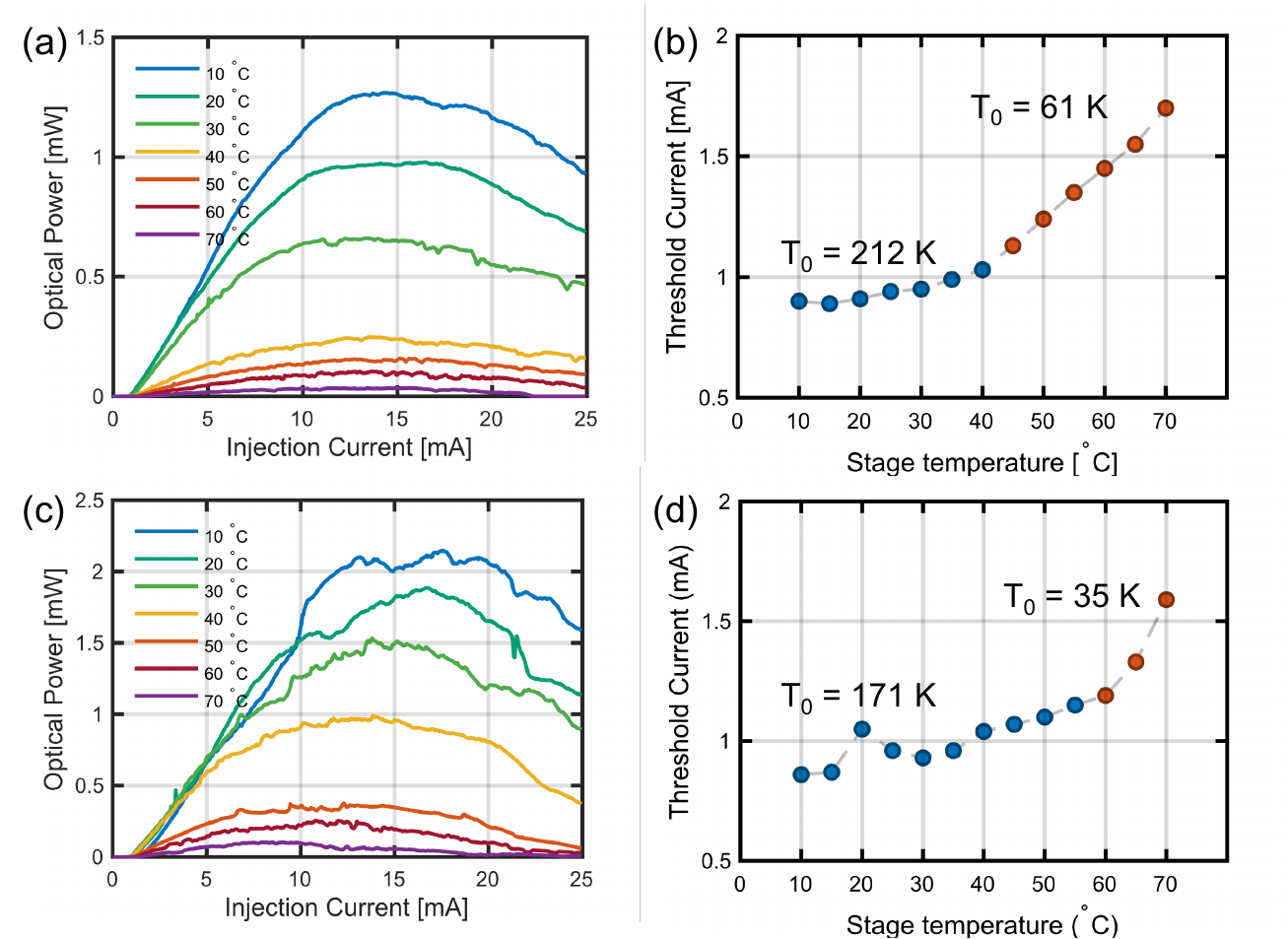}
\caption{Thermal performance of III-V/Si MRL design ($R_r = 25$~\textmu m, $w_\text{III-V} = 3.5$~\textmu m, $w_\text{si} = 1.0$~\textmu m, $w_\text{bus} = 0.8$~\textmu m, $\theta_\text{coupler} = 80°$) with measured (a) LI curves and (b) threshold current data along with characteristic temperature as a function of stage temperature. Thermal performance of III-V/Si MRL design ($R_r = 25$~\textmu m, $w_\text{III-V} = 5$~\textmu m, $w_\text{si} = 1.0$~\textmu m, $w_\text{bus} = 0.8$~\textmu m, $\theta_\text{coupler} = 80°$) with measured (c) LI curves and (d) threshold current data along with characteristic temperature as a function of stage temperature.}
\label{Fig12}
\end{figure}

It is generally known that threshold currents $I_\text{th}$ of QD lasers exhibit weak temperature dependence due to the discrete density of states and strong carrier localization inherent to QD active regions. In contrast to QW devices, where carriers can readily thermally redistribute into higher-energy states or escape from the active region, the 3-D confinement in quantum dots suppresses carrier leakage and reduces the impact of thermally activated non-radiative recombination processes. We investigated the thermal performance of two MRL designs with identical design parameters (indicated in Fig.~\ref{Fig12} caption) with the exception of the mesa width $w_\text{III-V}$. 

Fig.~\ref{Fig12}a--b illustrates thermal performance for $w_\text{III-V} = 3.5$~\textmu m, whereas Fig.~\ref{Fig12}c--d is for $w_\text{III-V} = 5.0$~\textmu m. Thermal rollover for the MRLs persists given the small QD active region volume compared to FP cavity configurations with larger volumes (2400~\textmu m$^3$) which exhibit minimal rollover~\cite{Kurczveil2016_OE}. The characteristic temperature $T_0$, an indicator of temperature sensitivity, can be determined by $T_0 = \left[\frac{d}{dT}\ln(I_\text{th})\right]^{-1}$ and is calculated in Fig.~\ref{Fig12}b,~d for the two MRL lasers with different mesa widths. For $w_\text{III-V} = 3.5$~\textmu m, a slight increase of $I_\text{th}$ between 10--40~°C yields $T_0 = 212$~K, while a lower $T_0 = 61$~K is obtained in the higher temperature range of 40--70~°C. For the wider mesa width of $w_\text{III-V} = 5.0$~\textmu m, $I_\text{th}$ remains weakly temperature dependent over a broader range of 10--60~°C, corresponding to $T_0 = 171$~K, before decreasing to $T_0 = 35$~K in the 60--70~°C range. The temperature partitioning of the wide-mesa design is similar to that reported for previously demonstrated hybrid QD microring lasers, while the low-temperature $T_0$ in our device is higher by about 90~K and the high-temperature $T_0$ is comparable~\cite{Zhang2019_Optica}. Although the narrow-mesa design exhibits higher extracted $T_0$ values within each fitted segment, it enters the high-temperature degradation regime at a much lower temperature of 40~°C, indicating poorer overall high-temperature robustness. This is likely due to the larger thermal resistance associated with the smaller active-region volume, together with stronger optical mode overlap with the etched sidewalls, which enhances surface recombination. Higher $T_0$ values can be realized by reducing Auger recombination as a result of $p$-type doping~\cite{Gerschutz2008_CLEO,Marko_JSTQE,Fathpour2004_APL}.

In addition, temperature increases can still degrade performance through homogeneous broadening of the quantum dot gain spectrum and increased carrier escape into the wetting layer or barrier states. The overall $T_0$ is still limited by strong self-heating associated with the 2~\textmu m BOX layer, which increases the thermal resistance and suppresses heat removal from the active region~\cite{Zhang2019_Optica}. Introducing thermal shunts that connect the hot region to the silicon substrate should therefore further improve the high-temperature performance. It should be noted that during high temperature measurement, a closely placed optical fiber will experience mechanical vibrations from inhomogeneous thermal expansion of the surrounding air. To mitigate these vibrations, the optical fiber was positioned further away from the grating couplers and will lead to reduced optical power coupling efficiency.

\subsection{High-Speed Measurements}
In addition to the demonstration of ultra-low threshold and high WPE operation, we investigated the intrinsic modulation bandwidth of the MRLs used in Fig.~\ref{Fig12}. We performed high speed, small-signal $S$-parameter characterization using an Agilent E8363B vector network analyzer (VNA) capable of 50 GHz bandwidth. The device is probed  using a signal/ground (SG) RF probe, with the sample placed on a temperature-controlled stage held at T = 25~°C. The measurements were performed on the same two devices used in Fig.~\ref{Fig12}: a narrow-mesa MRL ($R_r = 25$~\textmu m, $w_\text{III-V} = 3.5$~\textmu m, $w_\text{si} = 1.0$~\textmu m, $w_\text{bus} = 0.8$~\textmu m, $\theta_\text{coupler} = 80^\circ$) and a wide-mesa MRL ($R_r = 25$~\textmu m, $w_\text{III-V} = 5$~\textmu m, $w_\text{si} = 1.0$~\textmu m, $w_\text{bus} = 0.8$~\textmu m, $\theta_\text{coupler} = 80^\circ$). These two devices were chosen because their high slope efficiencies provide sufficient photoreceiver signal-to-noise ratio for reliable extraction of the relaxation oscillation frequency $f_r$ and damping rate $\gamma$ over a wide bias range (from 2.0 to 12.0~mA).

\begin{figure}[htbp]
\centering\includegraphics[width=13cm]{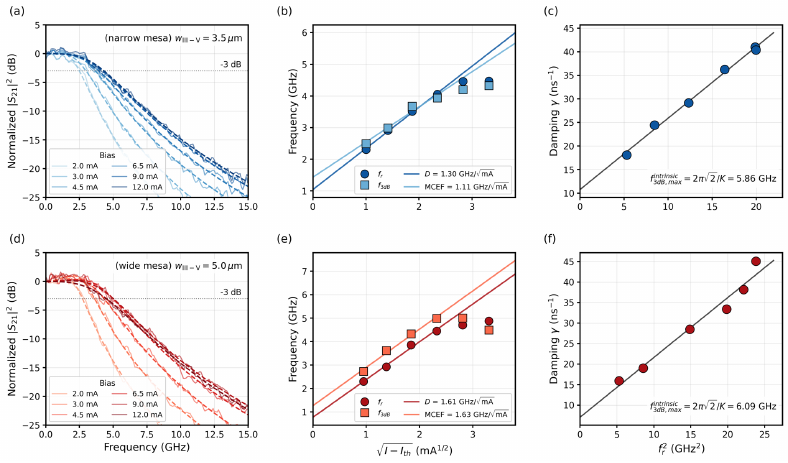}
\caption{Small-signal modulation response of III-V/Si MRL design ($R_r = 25$~\textmu m, $w_\text{III-V} = 3.5$~\textmu m, $w_\text{si} = 1.0$~\textmu m, $w_\text{bus} = 0.8$~\textmu m, $\theta_\text{coupler} = 80°$) and MRL design ($R_r = 25$~\textmu m, $w_\text{III-V} = 5$~\textmu m, $w_\text{si} = 1.0$~\textmu m, $w_\text{bus} = 0.8$~\textmu m, $\theta_\text{coupler} = 80°$) with (a,~d) measured normalized $|S_{21}|^2$ responses biased from 2.0 to 12.0~mA, where the fitting curves (dashed) are drawn using a three-pole fitting function, (b,~e) 3-dB bandwidth $f_\text{3dB}$ and relaxation oscillation frequency $f_r$ versus square root of bias current above threshold, with linear fits yielding the $D$-factor and modulation current efficiency factor (MCEF), and (c) damping rate $\gamma$ versus squared $f_r$ with linear fit $\gamma = K \cdot f_r^2 + \gamma_0$. The maximum 3-dB bandwidth limited by the $K$-factor, $f_\text{3dB,max}$, is 5.86~GHz and 6.09~GHz, respectively.}
\label{Fig13}
\end{figure}

\begin{figure}[htbp]
\centering\includegraphics[width=13cm]{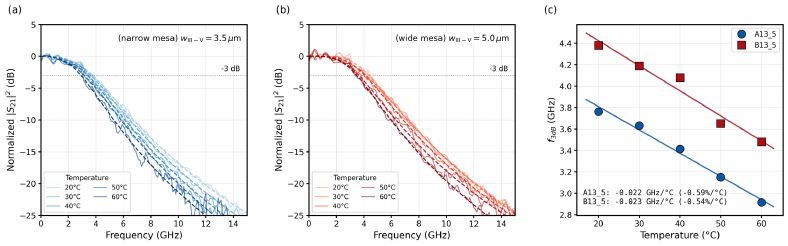}
\caption{Temperature dependence of small-signal modulation response of III-V/Si MRL design ($w_\text{III-V} = 3.5$~\textmu m) with (a) measured and fitted normalized $|S_{21}|^2$ responses at stage temperatures from 20~°C to 60~°C. Temperature dependence of small-signal modulation response of III-V/Si MRL design ($w_\text{III-V} = 5$~\textmu m) with (b) measured normalized $|S_{21}|^2$ responses at the same temperatures. (c) Extracted 3-dB bandwidth $f_\text{3dB}$ versus stage temperature for both devices with linear fits. Bias currents are fixed at 6.5~mA.}
\label{Fig14}
\end{figure}

The results are illustrated in Fig.~\ref{Fig13} where Fig.~\ref{Fig13}a,d show the normalized $|S_{21}|^2$ responses measured from 0.04 -- 15~GHz; the experimental responses (solid lines) are well fitted by the standard three-pole modulation transfer function (dashed lines)~\cite{Nagarajan1992_JQE}, from which $f_r$ and $\gamma$ are extracted at each bias and the 3-dB bandwidth $f_\text{3dB}$ is read directly from the measured data. Fig.~\ref{Fig13}b,e and Fig.~\ref{Fig13}c,f plot $f_r$ and $f_\text{3dB}$ versus the square root of $(I - I_\text{th})$ and $\gamma$ versus $f_r^2$, respectively, from which we extract the $D$-factor, the modulation current efficiency factor (MCEF) from the linear slopes, and the $K$-factor according to the relation~\cite{Inoue2018_OE}:
\begin{equation}
    \gamma = K \cdot f_r^2 + \gamma_0,
    \label{eq:damping}
\end{equation}
where $\gamma_0$ represents the damping offset and the corresponding intrinsic-damping-limited 3-dB bandwidth ceiling for each device. Linear fits in Fig.~\ref{Fig13}b,e and Fig.~\ref{Fig13}c,f yield $D$-factors of 1.30~GHz/mA$^{1/2}$ for the 3.5~\textmu m mesa device and 1.61~GHz/mA$^{1/2}$ for the 5~\textmu m mesa device, and MCEF values of 1.11~GHz/mA$^{1/2}$ and 1.63~GHz/mA$^{1/2}$, respectively --- a 24\% increase in $D$ and a 47\% increase in MCEF for the wider III-V mesa. The observed increase indicates that the differential gain $dg/dN$ is significantly higher in the wide-mesa device. The extracted $K$-factors are 1.52~ns and 1.46~ns; the slightly lower $K$ of the 5~\textmu m mesa device translates into a marginally higher intrinsic-damping-limited 3-dB bandwidth ceiling (6.09~GHz) compared to the 3.5~\textmu m mesa device (5.86~GHz). The maximum measured $f_\text{3dB}$ values of 4.32~GHz for the 3.5~\textmu m mesa device and 5.0~GHz for the 5~\textmu m mesa device occupy 73\% and 82\% of their respective intrinsic ceilings --- the 5~\textmu m device thus operates not only with a higher $K$-limited ceiling but also closer to it, as further indicated by the visible sublinear roll-off of $f_\text{3dB}$ at the highest bias points in Fig.~\ref{Fig13}e.

To assess the thermal robustness of the modulation response, $S$-parameter measurements were repeated at stage temperatures from 20~°C to 60~°C in 10~°C increments, with fixed bias currents of 6.5~mA for both devices. The normalized $|S_{21}|^2$ spectra in Fig.~\ref{Fig14}a,b exhibit a progressive roll-off with increasing temperature, well fitted by three-pole fits (dashed lines) across all measured temperatures. The extracted $f_\text{3dB}$ versus stage temperature is plotted in Fig.~\ref{Fig14}c, decreasing approximately linearly from 3.76~GHz at 20~°C to 2.92~GHz at 60~°C for the 3.5~\textmu m device, and from 4.38~GHz to 3.48~GHz for the 5~\textmu m device. Linear fits yield thermal degradation rates $df_\text{3dB}/dT = -0.022$~GHz/°C for the 3.5~\textmu m device and $-0.023$~GHz/°C for the 5~\textmu m device --- comparable in absolute terms. Normalized to the room-temperature bandwidth, however, the relative degradation rates are $-0.59$\%/°C for the 3.5~\textmu m device and $-0.54$\%/°C for the 5~\textmu m device, indicating that the 5~\textmu m mesa design exhibits a smaller fractional bandwidth penalty under thermal load.

\begin{figure}[htbp]
\centering\includegraphics[width=13cm]{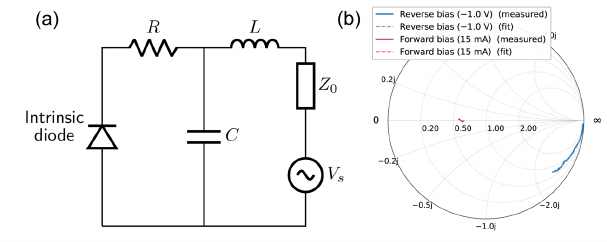}
\caption{(a) Equivalent circuit model used for the fitting. (b) Measured and fitted curves of reflection characteristics of the 5~\textmu m mesa device for reverse ($-1$~V) and forward (15~mA) biased conditions from 0.04 -- 5~GHz.}
\label{Fig15}
\end{figure}

We now turn to the small-signal impedance response to identify the parasitic elements that limit $f_\text{3dB}$ below its intrinsic ceiling. One-port $S_{11}$ measurements were performed on the 3.5~\textmu m device under both forward (15~mA) and reverse ($-1.0$~V) bias and fit to the equivalent circuit of Fig.~\ref{Fig15}a, comprising an intrinsic-diode branch in series with an internal resistance $R$, in parallel with a total capacitance $C$ that encompasses both the geometric pad capacitance and the bias-dependent diode junction capacitance, with the network in series with a parasitic inductance $L$ and the 50~$\Omega$ source impedance $Z_0$ \cite{Bowers1986_JQE}. The forward-bias measurement yields $R = 31.3$~$\Omega$, $L = 0.34$~nH, and a total capacitance $C_\text{total} = 0.355$~pF, while the reverse-bias measurement yields the pad capacitance $C_\text{pad} = 0.235$~pF. Compared to the device of Wan et al.~\cite{Wan2018_PR}, which reports $C_\text{pad} = 0.33$~pF on a similar III-V/Si quantum-dot ring laser, the value of 0.235~pF achieved here represents a ${\sim}30\%$ reduction attributable to the BCB planarization layer that increases the effective dielectric thickness between the metal contact pad and the underlying substrate. The corresponding RC cutoff frequency $f_\text{RC} = 1/(2\pi R C_\text{total}) = 14.3$~GHz lies well above the $K$-factor-limited intrinsic ceiling (5.86~GHz for the 3.5~\textmu m device), indicating that parasitic capacitance is not the dominant limiting factor on the measured $f_\text{3dB}$.

\section{Conclusion}
In this work, we present a systematic design space exploration and experimental validation of hybrid III-V/Si InAs/GaAs quantum dot micro-ring lasers targeting ultra-low threshold, high efficiency operation at 1.3~\textmu m for silicon photonic integration. Through multi-dimensional co-optimization of the resonator geometry, active region confinement, and coupling conditions, we identified the key trade-offs governing threshold current, slope efficiency, wall-plug efficiency, and thermal performance across a large fabricated design-of-experiment matrix.

On the design side, finite-difference eigenmode calculations revealed that the QD confinement factor $\Gamma_\text{5QD}$ is strongly dependent on both the silicon waveguide width $w_\text{si}$ and the ring radius $R_r$, with narrower waveguides and tighter bends increasing $\Gamma_\text{5QD}$ at the cost of reduced silicon modal confinement and elevated radiation loss. Three-dimensional FDTD simulations of the wrapped bus directional coupler demonstrated that extending the coupling angle $\theta_\text{coupler}$ provides a practical, monotonically tunable handle on the power coupling coefficient $\kappa$, enabling independent control of threshold current and slope efficiency within a single lithographic degree of freedom. Analytical modeling based on the full threshold and slope efficiency expressions, incorporating measured and vendor-supplied material parameters, confirmed that sub-milliamp thresholds are achievable for injection efficiencies $\eta_i > 0.75$ across a wide range of $\kappa$ values, and that reducing $R_r$ by 40\% can improve $I_\text{th}$ by 33--50\% at high injection efficiency.

Experimentally, devices from Groups A ($w_\text{III-V} = 3.5$~\textmu m) and B ($w_\text{III-V} = 5.0$~\textmu m) demonstrated average threshold currents of 1.43~mA and 1.34~mA, respectively, with select devices achieving record performance metrics among electrically driven hybrid III-V/Si micro-ring lasers: a minimum threshold current density of 108.9~A/cm$^2$ ($I_\text{th} = 0.77$~mA), a peak wall-plug efficiency of 9.78\%, a slope efficiency of 0.169~W/A, and output powers exceeding 2~mW. Model-experiment fitting across the coupling angle sweep yielded fitted injection efficiencies of $\eta_i = 0.50$ and $\eta_i = 0.31$ for the wide- and narrow-mesa devices, respectively, with the lower value for the narrow mesa attributed to enhanced surface recombination at the etched sidewalls.

Thermal characterization revealed characteristic temperatures of $T_0 = 212$~K (10--40~°C) and $T_0 = 171$~K (10--60~°C) for the narrow- and wide-mesa designs, respectively, representing the highest reported low-temperature $T_0$ values among hybrid QD micro-ring lasers and confirming the intrinsic thermal stability of the QD gain medium. The onset of accelerated threshold degradation at 40~°C for the narrow-mesa device underscores the importance of thermal management in compact resonator geometries, and thermal shunts connecting the active region to the silicon substrate are identified as a promising path toward improved high-temperature robustness. Small-signal modulation measurements yielded $K$-factor-limited 3-dB bandwidth ceilings of 5.86~GHz and 6.09~GHz for the narrow- and wide-mesa devices, with the wide-mesa device achieving a maximum measured $f_\text{3dB}$ of 5.0~GHz and operating at 82\% of its intrinsic ceiling, consistent with its higher differential gain. Parasitic analysis confirmed that the RC cutoff frequency of 14.3~GHz lies well above the intrinsic damping limit, establishing gain damping rather than parasitic capacitance as the dominant bandwidth constraint.

Collectively, these results demonstrate that systematic design optimization --- spanning confinement engineering, coupler geometry, and mesa scaling --- is essential for simultaneously achieving low threshold, high efficiency, and adequate modulation bandwidth in hybrid III-V/Si QD micro-ring lasers. The design guidelines and experimental benchmarks established here provide a quantitative foundation for next-generation photonically integrated light sources targeting DWDM data communications, co-packaged optics, and energy-efficient optical interconnect architectures.

\section{Back matter}

Back matter sections should be listed in the order Funding/Acknowledgment/Disclosures/Data Availability Statement/Supplemental Document section. An example of back matter with each of these sections included is shown below. The section titles should not follow the numbering scheme of the body of the paper. 

\begin{backmatter}
\bmsection{Funding}
Advanced Research Projects Agency-Energy (DE-AR0001039). 

\bmsection{Acknowledgment}
We thank the UCSB nanofabrication facilities. 

\bmsection{Disclosures}
The authors declare no conflicts of interest.

\bmsection{Data Availability Statement}
The data that support the findings of this study are not publicly available at this time but may be obtained from the authors upon reasonable request. 

\bmsection{Supplemental document}
A supplemental document must be called out in the back matter so that a link can be included. For example, “See Supplement 1 for supporting content.” Note that the Supplemental Document must also have a callout in the body of the paper.

\end{backmatter}

\section{References}
\label{sec:refs}

%%%%%%%%%%%%%%%%%%%%%%% References %%%%%%%%%%%%%%%%%%%%%%%%%

%%%%%%%%%% If using BibTeX:
\bibliography{sample}

%%%%%%%%%% If preparing manually:
% \begin{thebibliography}{1}
% \newcommand{\enquote}[1]{``#1''}

% \bibitem{Zhang:14}
% Y.~Zhang, S.~Qiao, L.~Sun, Q.~W. Shi, W.~Huang, L.~Li, and Z.~Yang,
%   \enquote{Photoinduced active terahertz metamaterials with nanostructured
%   vanadium dioxide film deposited by sol-gel method,}
%   {\protect\JournalTitle{Optics Express}} \textbf{22}, 11070--11078 (2014).

% \bibitem{Optica}
% {Optica}, \enquote{{Optica Publishing Group},}
%   \url{http://www.opg.optica.org}.

% \bibitem{FORSTER2007}
% P.~Forster, V.~Ramaswamy, P.~Artaxo, T.~Bernsten, R.~Betts, D.~Fahey,
%   J.~Haywood, J.~Lean, D.~Lowe, G.~Myhre, J.~Nganga, R.~Prinn, G.~Raga,
%   M.~Schulz, and R.~V. Dorland, \enquote{Changes in atmospheric consituents and
%   in radiative forcing,} in \enquote{Climate Change 2007: The Physical Science
%   Basis. Contribution of Working Group 1 to the Fourth Assesment Report of
%   Intergovernmental Panel on Climate Change,}  S.~Solomon, D.~Qin, M.~Manning,
%   Z.~Chen, M.~Marquis, K.~B. Averyt, M.~Tignor, and H.~L. Miler, eds.
%   (Cambridge University Press, 2007).

% \end{thebibliography}

\end{document}

\maketitle

\section{Threshold Current Extraction from LIV Curves}

The threshold current of the quantum-dot microring lasers was determined from the measured LI curves using the 2nd-derivative-based method [51-53]. Compared with direct inspection of the LI characteristics, this method provides a more stable means of locating the threshold region around the inflection point of the curve. Furthermore, a local parabolic interpolation near the maximum of the second derivative was employed to improve the extraction accuracy for discretely sampled data.

\subsubsection*{1.1\quad LIV Data Acquisition and Preprocessing}

The LI characteristics were measured by sweeping the gain current with a step size of 0.05~mA. Because the threshold current of the devices is generally below 4~mA, only the 0--4~mA portion of each LI curve was included in the threshold current analysis. Limiting the analysis to this low-current region reduces the risk of false threshold identification arising from complex mode switching, mode hopping, and abrupt power discontinuities at higher injection currents.

\subsubsection*{1.2\quad Smoothing the Data Points}

\begin{figure}[htbp]
\centering\includegraphics[width=13cm]{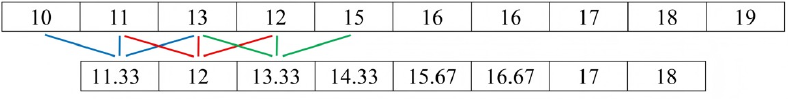}
\caption{Illustration of the moving-window average used for smoothing. A 3-point sliding window is shown here for simplicity, where the local average of three consecutive data points is assigned to the smoothed sequence.}
\label{Fig1_supp}
\end{figure}

Since numerical differentiation is highly sensitive to high-frequency fluctuations in the measured LI data, direct calculation of the derivatives may introduce artificial sharp peaks in the 2nd derivative. To mitigate this issue, a moving-window averaging method was employed for data smoothing. Specifically, the smoothed data points were generated by applying a moving-window average along the measured dataset. Each smoothing window contained 10 consecutive data points, and the average value within the window was taken as one element of the smoothed data sequence. A schematic illustration of this moving-average procedure is shown in Fig.~\ref{Fig1_supp}.

The smoothing was first performed on the raw LI curve and subsequently on the first-derivative curve, to minimize noise amplification during 2nd-derivative extraction. An example of the resulting smoothed LI and 1st-derivative curves is shown in Fig.~\ref{Fig2_supp}a,b.

\subsubsection*{1.3\quad 1$^{st}$ and 2$^{nd}$ Derivative Calculation}

Because the LI characteristics were recorded as discrete data points, numerical derivatives were evaluated using finite differences. The resulting raw and smoothed first-derivative curves are shown in Fig.~\ref{Fig2_supp}b. For the smoothed LI dataset $\{I_k,\, \tilde{P}_k\}$, the first derivative was calculated as
\begin{equation}
    \left(\frac{dP}{dI}\right)_k \approx \frac{\tilde{P}_{k+1} - \tilde{P}_k}{I_{k+1} - I_k},
    \label{eq:first_deriv}
\end{equation}
with the associated current coordinate assigned to the midpoint of two adjacent sampling points,
\begin{equation}
    I_k^{(1)} = \frac{I_{k+1} + I_k}{2}.
    \label{eq:midpoint}
\end{equation}
The 1st-derivative data were then smoothed again using the same moving-average procedure, after which the 2nd derivative was calculated in the same manner:
\begin{equation}
    \left(\frac{d^2P}{dI^2}\right)_k \approx \frac{\widetilde{(dP/dI)}_{k+1} - \widetilde{(dP/dI)}_k}{I_{k+1}^{(1)} - I_k^{(1)}}.
    \label{eq:second_deriv}
\end{equation}
The current coordinate of each 2nd-derivative point was likewise assigned to the midpoint between two neighboring 1st-derivative coordinates.

\subsubsection*{1.4\quad Parabolic Interpolation Around the Second-Derivative Peak}

\begin{figure}[htbp]
\centering\includegraphics[width=13cm]{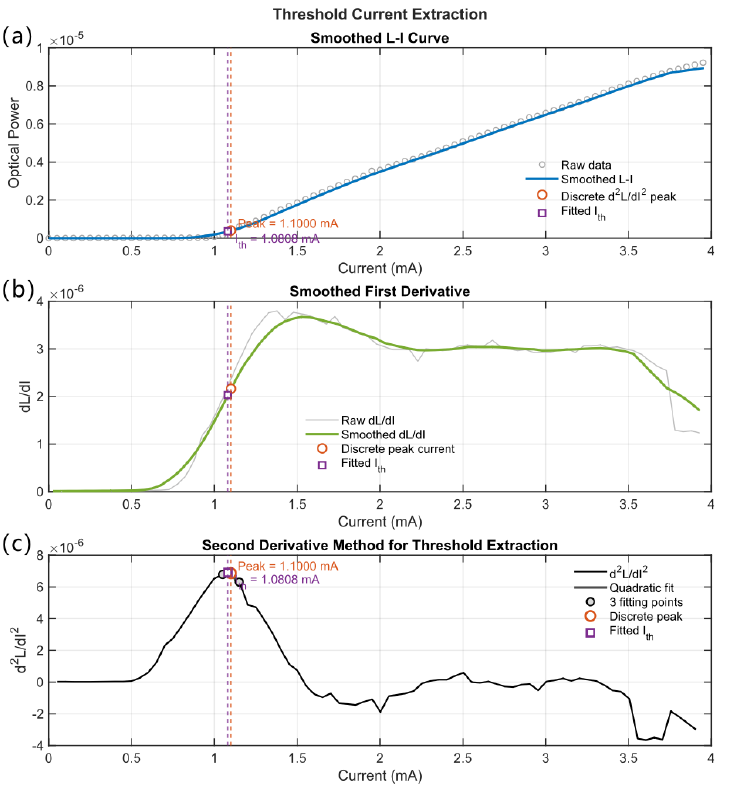}
\caption{Example of threshold-current extraction using the 2nd-derivative method. (a) The raw and smoothed LI curve, (b) the raw and smoothed 1st-derivative curve, (c) the 2nd-derivative curve used for threshold determination. The discrete peak of the 2nd derivative and the threshold current $I_\text{th}$ obtained from local quadratic fitting are both marked. The fitted $I_\text{th}$ is also projected onto the smoothed LI and 1st-derivative curves for visual comparison.}
\label{Fig2_supp}
\end{figure}

The threshold current $I_\text{th}$ was extracted from the 2nd-derivative curve in the selected current window. As shown in Fig.~\ref{Fig2_supp}c, the discrete maximum of $d^2P/dI^2$ was first located to identify the approximate threshold region. Subsequently, the peak point together with its two neighboring data points was fitted by a quadratic function,
\begin{equation}
    y = aI^2 + bI + c.
    \label{eq:parabola}
\end{equation}
The vertex of this parabola was then used to determine the final threshold current according to
\begin{equation}
    I_\text{th} = -\frac{b}{2a}.
    \label{eq:Ith_parabola}
\end{equation}
Compared with directly assigning the threshold current to the discrete second-derivative maximum, this local parabolic fitting reduces the error introduced by finite current steps and yields a more precise estimate of the threshold position.

\section{Extraction of characteristic temperature T$_0$ }

The characteristic temperature $T_0$ was extracted from the temperature dependence of the threshold current using the standard empirical relation for semiconductor lasers,
\begin{equation}
    I_\text{th}(T) = I_0 \exp\!\left(\frac{T}{T_0}\right),
    \label{eq:T0}
\end{equation}
where $I_\text{th}$ is the threshold current at temperature $T$, and $I_0$ is a fitting constant. Taking the natural logarithm gives
\begin{equation}
    \ln\!\left(I_\text{th}(T)\right) = \ln(I_0) + \frac{T}{T_0},
    \label{eq:lnIth}
\end{equation}
which converts the exponential dependence into a linear form. Therefore, $T_0$ can be obtained from the inverse of the slope of a linear fit to $\ln(I_\text{th})$ versus $T$,
\begin{equation}
    T_0 = \left(\frac{d\ln(I_\text{th})}{dT}\right)^{-1}.
    \label{eq:T0_slope}
\end{equation}
This form is used throughout this work to quantify the temperature sensitivity of the threshold current.

\begin{figure}[htbp]
\centering\includegraphics[width=13cm]{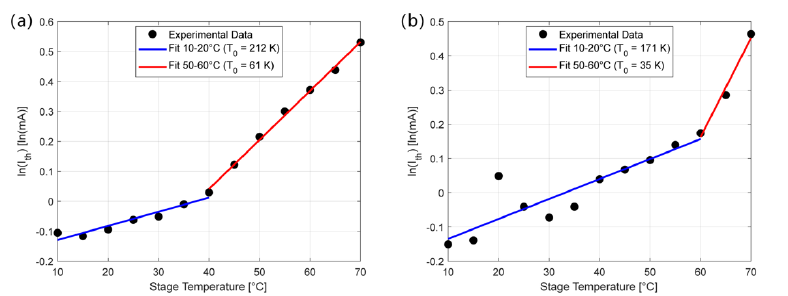}
\caption{$\ln(I_\text{th})$ versus stage temperature for two III-V/Si QD microring lasers with different mesa widths: (a) $R_r = 25$~\textmu m, $w_\text{III-V} = 3.5$~\textmu m, $w_\text{si} = 1.0$~\textmu m, $w_\text{bus} = 0.8$~\textmu m, $\theta_\text{coupler} = 80°$ and (b) $R_r = 25$~\textmu m, $w_\text{III-V} = 5$~\textmu m, $w_\text{si} = 1.0$~\textmu m, $w_\text{bus} = 0.8$~\textmu m, $\theta_\text{coupler} = 80°$. Segmented linear fits were used to extract the characteristic temperature $T_0$. The extracted values are 212~K (10--40~°C) and 61~K (40--70~°C) for (a), and 171~K (10--60~°C) and 35~K (60--70~°C) for (b).}
\label{Fig3_supp}
\end{figure}

To extract $T_0$ from the measured data, the threshold current $I_\text{th}$ was first determined at each stage temperature using the method described above. The natural logarithm of the extracted threshold current, $\ln(I_\text{th})$, was then plotted as a function of temperature, and linear fitting was applied to the resulting $\ln(I_\text{th})$--$T$ relation, as shown in Fig.~\ref{Fig3_supp}. Since the data were not well described by a single straight line over the full temperature range, segmented linear fitting was employed. For the narrow-mesa device, two fitting regions were used: 10--40~°C and 40--70~°C. For the wide-mesa device, the fitting regions were 10--60~°C and 60--70~°C. These fitting windows were chosen based on the clear change in slope observed in the experimental data, which indicates the onset of a distinct high-temperature degradation regime. The slope extracted from each segment was then used to determine the corresponding $T_0$ value for each device.

%Manual citation list
%\begin{thebibliography}{1}
%\bibitem{Zhang:14}
%Y.~Zhang, S.~Qiao, L.~Sun, Q.~W. Shi, W.~Huang, %L.~Li, and Z.~Yang,
 % \enquote{Photoinduced active terahertz metamaterials with nanostructured
  %vanadium dioxide film deposited by sol-gel method,} Opt. Express \textbf{22},
  %11070--11078 (2014).
%\end{thebibliography}